\documentclass{JHEP3}
\usepackage{amsmath, amssymb, stmaryrd}

\def\ba{\begin{align}}
\def\ea{\end{align}}
\def\be{\begin{equation}}
\def\ee{\end{equation}}
\def\nn{\nonumber}
\def\exd{{\rm d}}
\def\pd{\partial}
\makeatletter
\def\x@arrow{\DOTSB\Relbar}
\def\xlongequalsignfill@{\arrowfill@\x@arrow\Relbar\x@arrow}
\newcommand{\xlongequal}[2]{%
    \ext@arrow 0099\xlongequalsignfill@{#1}{#2}}
\makeatother

\newcommand{\roughly}[1]{\mathrel{\raise.3ex\hbox{$#1$\kern-0.85em
\lower1ex\hbox{$\sim$}}}}

\def\endignore{}
\def\ignore #1\endignore{} 

\def\be{\begin{equation}}
\def\beq\begin{equation}
\def\ee{\end{equation}}
\def\bea{\begin{eqnarray}}
\def\eea{\end{eqnarray}}

\def\nn{\nonumber}

\def\pref#1{(\ref{#1})}

\def\beq{\begin{equation}}
\def\eeq{\end{equation}}
\def\beqa{\begin{eqnarray}}
\def\eeqa{\end{eqnarray}}

\def\cB{{\cal B}}

\def\cF{{\cal F}}
\def\cG{{\cal G}}

\def\cJ{{\cal J}}

\def\cL{{\cal L}}
\def\cM{{\cal M}}

\def\cR{{\cal R}}

\def\cT{{\cal T}}
\def\cU{{\cal U}}

\def\cW{{\cal W}}

\def\ssA{{\scriptscriptstyle A}}
\def\ssB{{\scriptscriptstyle B}}
\def\ssC{{\scriptscriptstyle C}}
\def\ssD{{\scriptscriptstyle D}}

\def\ssM{{\scriptscriptstyle M}}
\def\ssN{{\scriptscriptstyle N}}
\def\ssP{{\scriptscriptstyle P}}
\def\ssQ{{\scriptscriptstyle Q}}

\def\GH{{\scriptscriptstyle GH}}

\def\ansatz{{\it ansatz}}

\newcommand{\bmat}{\left(\begin{array}}
\newcommand{\emat}{\end{array}\right)}

\def\-{\hphantom{-}}

\def\s2{\frac{1}{2}}

\def\IF{\relax{\rm I\kern-.18em F}}
\def\II{\relax{\rm I\kern-.18em I}}
\def\IP{\relax{\rm I\kern-.18em P}}
\def\IC{\relax{\rm I\kern-.48em C}}
\def\IR{\relax{\rm I\kern-.18em R}}
\def\IK{\relax{\rm I\kern-.20em K}}
\def\IM{\relax{\rm I\kern-.25em M}}

\def\Dsl{\,\raise.15ex\hbox{/}\mkern-13.5mu D}
\def \one{\relax{\rm 1\kern-.26em I}}

\def\exd{{\rm d}}

\author{Allan Bayntun,${}^1$ C.P.~Burgess${}^{1,2}$
and Leo van Nierop${}^1$ \\

${}^1$ Department of Physics \& Astronomy\\
$\phantom{D}$ McMaster University, Hamilton ON, Canada, L8S 4M1. \\
${}^2$ Perimeter Institute for Theoretical Physics\\
 $\phantom{D}$ 31 Caroline St. N,
 Waterloo ON, Canada, N2L 2Y5. \\}

\title{Codimension-2 Brane-Bulk Matching:\\
Examples from Six and Ten Dimensions}

\abstract{Experience with Randall-Sundrum models teaches the
importance of following how branes back-react onto the bulk
geometry, since this can dramatically affect the system's
low-energy properties. Yet the practical use of this observation
for model building is so far mostly restricted to branes having
only one transverse dimension (codimension-1) in the bulk space,
since this is where tools for following back-reaction are
well-developed. This is likely a serious limitation since
experience also tells us that one dimension is rarely
representative of what happens in higher dimensions. We here
summarize recent progress on developing the matching conditions
that describe how codimension-2 branes couple to bulk metric,
gauge and scalar fields. These matching conditions are then
applied to three situations: $D7$-branes in F-theory
compactifications of 10D Type IIB string vacua; 3-branes coupled
to bulk axions in unwarped and non-supersymmetric 6D systems; and
3-branes coupled to chiral, gauged 6D supergravity. For each it is
shown how the resulting brane-bulk dynamics is reproduced by the
scalar potential for the low-energy moduli in the dimensionally
reduced, on-brane effective theory. For 6D supergravity we show
that the only 4D-maximally symmetric bulk geometries supported by
positive-tension branes are flat. }

\begin{document}

\section{Introduction}

Space-filling branes, situated around extra dimensions, provide a
remarkable framework for approaching phenomenological problems.
Besides being well-motivated --- for instance arising very
naturally within string theory --- branes lead to novel kinds of
low-energy physics that can cut to the core of many of the
naturalness issues that currently plague particle physics and
cosmology.

The realization that not all particles need `see' the same number
of dimensions (because brane-bound particles are trapped to move
only along the branes) is the first type of brane-related insight
to have made a major impact on physics, leading to the recognition
that the scale of gravity can be much smaller than the Planck
scale \cite{braneworld}. A second major revelation came with the
realization that the back-reaction of branes on their environment
can strongly influence their low-energy properties, such as by
providing deep gravitational potential wells within the extra
dimensions that redshift the energy of those branes that live
within them \cite{RS}.

Although branes can in principle have a great variety of
dimensions, almost all of the detailed exploration of brane-bulk
back-reaction is specialized to the case of codimension-1 branes:
{\em i.e.} those branes that span just one dimension less than the
dimension of the full spacetime. This is partially because tools
for describing how branes back-react on their surroundings are
only well-developed for codimension-1 surfaces, since in this case
the problem can be expressed in terms of the Israel junction
conditions \cite{IJC}. This restriction to codimension-1 objects
is potentially very limiting because the special nature of
kinematics in one dimension makes it unlikely that back-reaction
for codimension-1 branes is representative of back-reaction for
branes with higher codimension.

The main obstacle to understanding how properties of
higher-codimension branes are related to the bulk geometries they
source is the fact that these bulk geometries typically diverge at
the position of their sources. (The most familiar example of this
for a codimension-3 object is the divergence of the Coulomb
potential of a nucleus evaluated at the nuclear position.) It is
one of the special features of codimension-1 objects that the bulk
fields they source typically do not diverge at their positions.
They instead cause discontinuities of derivatives across their
surfaces, whose properties are captured by the Israel junction
conditions.

The next-simplest case consists of codimension-2 objects, whose
back-reaction is complicated enough to allow the possibility of
bulk fields diverging at the positions of the sources. Although
bulk fields {\em can} diverge for codimension-2 sources, they {\em
needn't} do so in time-independent situations. (For instance, they
can instead give rise to conical singularities, such as for cosmic
strings in 4D spacetime \cite{cosmicstringbr}. When bulk fields do
not diverge the relation between bulk and brane properties is
easier to formulate, and so better studied \cite{conical}.) The
potential for divergent bulk configurations makes codimension-2
branes more representative of systems with more generic
codimension than are codimension-1 branes. But dynamics in two
dimensions is still simple enough to allow explicit closed-form
solutions to be known for the bulk configurations sourced by
codimension-2 branes, allowing a detailed study of their
properties.

Tools for describing how bulk fields respond to the properties of
source branes were recently developed in the general case,
including where the bulk fields diverge
\cite{uvcaps,cod2matching,otheruvcaps}, opening up the properties
of codimension-2 branes for phenomenological exploration. These
tools
--- summarized (and slightly generalized) in \S2\ below for a
fairly general class of scalar-tensor-Maxwell theories in $n$
extra dimensions --- boil down to a set of matching conditions
that relate the near-brane limit of the radial derivatives of the
bulk fields to the action for the brane in question.

In \S3\ we apply these tools to three kinds of examples: compact
geometries sourced by D7 branes in F-theory compactifications of
10D Type IIB supergravity; 3-branes coupled to a bulk axion within
unwarped, non-supersymmetric 6D scalar/Maxwell/Einstein theory;
and 3-branes coupled to 6D chiral gauged supergravity. We draw the
following lessons from these comparisons:
\begin{itemize}
\item F-theory compactifications \cite{ftheory} of 10D Type IIB
supergravity sourced by D7-branes serve as a reality check, since
string theory tells us the detailed form of both the brane and
bulk actions \cite{polchinski}, and explicit solutions are known
for the transverse spacetimes that are sourced by these branes
\cite{scs}. We verify the codimension-2 brane/bulk matching
conditions by checking that the asymptotic forms for the solutions
are related to the known brane actions in the prescribed way.
\item In 6D axion-Maxwell-Einstein theory, flux-compactified
solutions are known for the bulk that interpolates between two
3-branes, and these are simple enough to allow the explicit
calculation of how branes contribute to the low-energy axion
potential \cite{6dnonsusy}. From the perspective of six dimensions
the resulting axion stabilization arises through the requirement
that both branes be consistent in their demands on the bulk. We
show that the stabilized value agrees precisely with the result of
minimizing the low-energy axion potential as seen by an observer
who has integrated out the extra dimensions below the Kaluza-Klein
(KK) scale. We also show how this potential gives the same value
for the curvature of the maximally symmetric on-brane geometry as
is calculated from the higher-dimensional field equations.
\item Stable flux compactifications are also known for 6D chiral
gauged supergravity \cite{NS}, having up to two singularities that
represent the positions of two source branes \cite{6dsusy}. These
solutions are known in explicit closed form for the most general
solutions having a flat on-brane geometry and axial symmetry in
the bulk; and in a slightly more implicit form for solutions with
de Sitter or anti-de Sitter on-brane geometry. In this case we use
the matching conditions to show that the only bulk configurations
that can be supported by positive-tension branes have flat induced
on-brane geometries, with (possibly warped) bulk geometries with
nonsingular limits as the source branes are approached. We also
show how geometries that diverge at the brane positions can arise
from specific kinds of negative-tension branes, while no maximally
symmetric solutions exist at all for many kinds of brane sources
(presumably corresponding to time-dependent runaway bulk
geometries, such as those considered in \cite{scalingpaper}).
\end{itemize}

\S4\ briefly summarizes some of the implications of these results.

\section{The Bulk-Brane system}

We start by describing the brane-bulk framework within which we
work. This starts with a statement of the scalar-metric-Maxwell
system whose equations we use, followed by a statement of how the
near-brane boundary conditions of the bulk fields are related to
the action of the branes which are their source. Finally we
describe the contribution of each brane to the low-energy scalar
potential that is valid over distances much longer than the size
of the extra dimensions, and identify a constraint which allows a
simple description of this contribution given the properties of
the brane tension.

\subsection{The bulk}

The starting point is the statement of the equations of motion
that govern the bulk.

\subsubsection*{General formulation}

We assume the following action for the $n$-dimensional bulk
physics, describing a general scalar-tensor theory coupled to a
Maxwell field,\footnote{Our metric is mostly plus, with Weinberg's
curvature conventions \cite{Wbg}, which differ from those of MTW
\cite{MTW} only by an overall sign in the definition of the
Riemann tensor.}
\begin{equation}
 S = \int_\cM \exd^{n} x \; \cL_\ssB +
 \int_{\partial \cM} \exd^{n-1} x
 \; \cL_{\GH}
\end{equation}
where
\begin{equation}
 \cL_\ssB = -  \sqrt{-g} \;
 \left\{ \frac{1}{2\kappa^2} \, g^{\ssM\ssN}\Bigl[
 \mathcal{R}_{\ssM\ssN}
 + \mathcal G_{\ssA \ssB}(\phi) \,\partial_\ssM \phi^\ssA
 \partial_\ssN \phi^\ssB \Bigr] + \frac14 \, f(\phi)
 \, F_{\ssM \ssN} F^{\ssM \ssN} + V(\phi) \right\} \,,
\end{equation}
and the Gibbons-Hawking lagrangian \cite{GH} is
\begin{equation}
 \cL_{\GH} = \frac{1}{\kappa^2}
 \sqrt{-\hat\gamma} \; K \,,
\end{equation}
and is required in the presence of boundaries in order to make the
Einstein action well posed. Here $F = \exd A$ is the field
strength of the Maxwell field, $\mathcal R$ is the Ricci scalar
for the 6D spacetime metric, $g_{\ssM \ssN}$, and $\mathcal
G_{\ssA \ssB}$ is the metric of the target space within which the
scalar fields, $\phi^\ssA$, $A = 1,\dots,N$, take values.
$\hat\gamma_{ij} = g_{\ssM \ssN} \, \partial_i x^\ssM \partial_j
x^\ssN$ is the induced metric, and $K$ is the trace, $\hat
\gamma^{ij} K_{ij}$, of the extrinsic curvature, of the boundary
surface, $\partial \cM$.

This bulk action is chosen to be general enough to include the
bosonic part of the supersymmetric theories of interest. Its field
equations are
\begin{align}
 \frac{1}{2\kappa^2} \left( \mathcal{R}_{\ssM\ssN}
 + \mathcal G_{\ssA \ssB} \, \partial_\ssM \phi^\ssA
 \partial_\ssN \phi^\ssB \right) +
 \frac{f}{2} \,  \, F_\ssM^{\ \ \ssP} F_{\ssN \ssP}
 + \frac{1}{n-2} \left[V - \frac{f}{4} \, F_{\ssP\ssQ}
 F^{\ssP\ssQ} \right] \,  g_{\ssM\ssN} &= 0,\\
 \mathcal G_{\ssA\ssB} \Box \,\phi^\ssB  - \kappa^2
 \left[ \frac{\pd V}{\partial\phi^\ssA}
 + \frac14 \, \frac{\pd
 f}{\pd \phi^\ssA} \,
 F_{\ssM\ssN} F^{\ssM\ssN} \right]&=0 \,,
\end{align}
and
\begin{equation}
 \nabla_\ssM \left( f F^{\ssM\ssN} \right) =0 \,,
\end{equation}
where
\begin{equation}
 \Box \,\phi^\ssA := g^{\ssM \ssN} \Bigl[ \nabla_\ssM \partial_\ssN
 \phi^\ssA + \Gamma^\ssA_{\ssB\ssC}(\phi) \partial_\ssM \phi^\ssB
 \partial_\ssN \phi^\ssC \Bigr] \,,
\end{equation}
with $\Gamma^\ssA_{\ssB\ssC}(\phi)$ being the Christoffel
connection built from the metric $\cG_{\ssA\ssB}$.

\subsubsection*{Metric {\em ans\"atze}}

Our interest is in configurations whose geometries are maximally
symmetric in the brane directions, for which it is convenient to
specialize to the metric
\begin{eqnarray} \label{metricansatz0}
 \exd s^2 = g_{\ssM\ssN} \, \exd x^\ssM \exd x^\ssN
 &=& e^{2W} \hat g_{\mu\nu} \, \exd x^\mu \exd x^\nu
 + g_{mn} \, \exd x^m \exd x^n \nn\\
 &=& e^{2W} \hat g_{\mu\nu} \, \exd x^\mu \exd x^\nu
 + e^{2C} \, \exd z \,\exd \overline z \,,
\end{eqnarray}
where $\hat g_{\mu\nu} (x)$ denotes a maximally symmetric
$(n-2)$-dimensional metric. The coordinates are $x^\ssM = \{
x^\mu, x^m \}$, with $x^\mu$, $\mu = 0, \dots, n-3$ labelling the
brane directions, and $m = n-2, n-1$ (or $z = x^{n-2} + i
x^{n-1}$) being coordinates for the two dimensions transverse to
the branes. The functions $W$ and $C$ are generally singular at
the positions of any source branes. For instance, if $e^{C} =
(\ell/r)^a$ for $r^2 = |z|^2$, then the proper distance becomes
$\rho = [\ell/(1-a)] (\ell/r)^{a-1}$ and $e^B = \ell
(\ell/r)^{a-1} = (1-a) \rho$, showing that the metric in this case
has a conical singularity at $r = \rho = 0$, with defect angle
$\delta = 2\pi a$.

For some applications, particularly very near a brane, it is
useful to further specialize to the most general \ansatz\
consistent with cylindrical symmetry in the two transverse
dimensions, $\{ x^m , m = n-2,n-1 \}$. This leads to the following
metric:
\begin{align} \label{metricansatz}
 \exd s^2 &= \exd\rho^2 + e^{2B} \exd\theta^2 + e^{2W}
 \hat g_{\mu\nu} \, \exd x^\mu \exd x^\nu \\
 &= e^{2C} \Bigl( \exd r^2 + r^2 \exd\theta^2
 \Bigr) + e^{2W}
 \hat g_{\mu\nu} \, \exd x^\mu \exd x^\nu \nn
\end{align}
where $\theta$ labels the direction of cylindrical symmetry, and
the functions $B = B(\rho)$ and $W = W(\rho)$ depend on the proper
distance, $\rho$, only --- or $C = C(r)$ is a function only of
$r$.

The bulk scalars are similarly just functions of $\rho$,
$\phi^\ssA = \phi^\ssA(\rho)$, and a gauge can be chosen to that
the only nonzero component for the Maxwell field is $A_\ssM =
A_\theta(\rho) \, \delta^\theta_\ssM$, and so
\begin{align}
 F_{\rho\theta}=-F_{\theta\rho} = A_\theta' \,,
\end{align}
where the prime denotes differentiation with respect to $\rho$.

The Einstein equations subject to this \ansatz\ reduce to
\begin{align}
 \frac1{n-2} \, e^{-2W} \, \hat R + W'' + (n-2) (W')^2
 + W' B' - \frac{1}{n-2} \kappa^2\, e^{-2B} \, f
 (A_{\theta}')^2 + \frac{2\kappa^2V}{n-2} &=0
 \text{\ \ ($\mu\nu$)} \\
 B'' + (B')^2 + (n-2) W' B' + \frac{n-3}{n-2}\kappa^2 \,
 e^{-2B} \, f (A_{\theta}')^2 + \frac{2\kappa^2V}{n-2}
 &=0 \text{\ \ ($\theta\theta$)} \\
 (n-2) \left[ W'' + (W')^2 \right] + B'' + (B')^2
 + \mathcal G_{\ssA\ssB} {\phi^\ssA}' {\phi^\ssB}'
 + \frac{n-3}{n-2}\kappa^2 \, e^{-2B} \, f (A_{\theta}')^2
 + \frac{2\kappa^2V}{n-2} &=0 \text{\ \ ($\rho\rho$)} \,,
\end{align}
while the dilaton and Maxwell equations become
\begin{equation}
 e^{-B - 4W}
 \Bigl( e^{B + 4W} \, \cG_{\ssA\ssB} {\phi^\ssB}' \Bigr)'
 + \cG_{\ssA\ssB} \Gamma^\ssB_{\ssC\ssD} \, {\phi^\ssC}'
 \, {\phi^\ssD}' - \kappa^2  \left[ \frac{\partial V}{\partial
 \phi^\ssA} + \frac14 \, \frac{\partial f}{\partial \phi^\ssA}
 \, e^{-2B} ( A_\theta')^2 \right] = 0 \,,
\end{equation}
and
\begin{equation}
 \Bigl( e^{-B + 4W} f \, A_\theta' \Bigr)' = 0 \,.
\end{equation}

\subsection{Boundary conditions for codimension-2 branes}

\subsubsection*{General formulation}

Suppose an $(n-2)$-dimensional, space-filling, codimension-2 brane
is located at a position, $x^m = x^m_b$, within the 2 extra
dimensions, with brane action
\begin{equation} \label{cod2braneaction}
 S_b = - \int_{x_b} \exd^{n-2} x \; \sqrt{-\gamma} \; \Bigl[
 L_b\left( \phi^\ssA, A_\theta, g_{\theta\theta} \right)
 + \cdots \Bigr] \,,
\end{equation}
where $L_b$ denotes the brane lagrangian, which is potentially a
function of the bulk scalars, $\phi^\ssA$, and the tangential
components of the bulk Maxwell field and metric, $A_\ssM$ and
$g_{\ssM\ssN}$, but not their derivatives. (Ellipses denote the
possible subdominant, higher-derivative effective interactions
that can also be present.) We imagine the geometry surrounding the
brane to be given by the axisymmetric {\it ansatz} of
eq.~\pref{metricansatz}, with the brane located at $\rho = 0$, so
$\theta$ denotes the angular direction about its position. Because
our interest is in maximally symmetric solutions along the brane
directions we do not entertain a dependence of $T_b$ on any
components of $A_\ssM$ and $g_{\ssM\ssN}$ apart from $A_\theta$
and $g_{\theta\theta}$.

The induced metric on the brane is $\gamma_{\mu\nu} = g_{\ssM\ssN}
\, \partial_\mu x^\ssM \, \partial_\nu x^\ssN = e^{2W} \hat
g_{\mu\nu}$. Because of the warp factor appearing in this metric,
for later purposes it is convenient to define the `warped'
tension, $T_b$, by $T_b = e^{(n-2) W} L_b$, so that the brane
action becomes
\begin{equation}
 S_b = - \int_{x_b} \exd^{n-2} x \; \sqrt{-\hat{g}} \; \Bigl[
 T_b(\phi,A_\theta,g_{\theta\theta},W) + \cdots \Bigr] \,.
\end{equation}

The back-reaction of such a brane onto the bulk geometry dictates
the asymptotic near-brane behaviour of the bulk fields
nearby,\footnote{A familiar example of this from electrostatics is
the $1/\rho$ dependence of the Coulomb potential that occurs in
the immediate vicinity of a point charge situated at $\rho=0$.}
through codimension-2 matching conditions that generalize
\cite{uvcaps,cod2matching,otheruvcaps} the more familiar ones that
are encountered for codimension-1 branes. For the bulk scalars
these state
\begin{equation}
 \lim_{\rho \to 0} \oint_{x_b} \exd \theta \;
 \left[ \frac{1}{\kappa^2} \sqrt{-g} \, \mathcal G_{\ssA \ssB}
 \partial_\rho\phi^\ssB \right]
 =  - \frac{\delta S_b}{\delta \phi^\ssA} \,,
\end{equation}
where the integration is about a small circle of proper radius
$\rho$ encircling the brane position, $x_b$, which is taken to be
situated at $\rho = 0$. Similarly, the Maxwell matching condition
is
\begin{equation}
  \lim_{\rho \to 0} \oint_{x_b} \exd \theta \;
   \Bigl[ \sqrt{-g} \; f \,
  F^{\rho\ssM} \Bigr]
  = - \frac{\delta S_b}{\delta A_\ssM} \,,
\end{equation}
Finally, the metric matching condition is
\begin{equation}
   \lim_{\rho \to 0} \oint_{x_b} \exd \theta \;
    \left[ \frac{1}{2\kappa^2} \, \sqrt{-g} \;
   \left( K^{ij} - K g^{ij} \right) - \hbox{(flat)}\right] = -
   \; \frac{\delta S_b}{\delta g_{ij}} \,,
\end{equation}
where $K_{ij}$ is the extrinsic curvature of the fixed-$\rho$
surface, for which the local coordinates are those appropriate for
surfaces of constant $\rho$: $\{ x^i , i = 0,1,\cdots,n-2 \}$.
Here `flat' denotes the same result evaluated near the origin of a
space for which the brane location $\rho = 0$ is nonsingular.

\subsubsection*{Axially symmetric ansatz}

Specialized to the \ansatz\ of eq.~\pref{metricansatz} the
scalar-field matching condition becomes
\begin{equation} \label{scalarmatching}
 \left[ \frac{2\pi}{\kappa^2} \, e^{B + (n-2)W} \sqrt{- \hat g}
 \, \mathcal G_{\ssA \ssB} \, {\phi^\ssB}' \right]_{x_b}
 = \frac{\pd }{\pd\phi^\ssA} \Bigl[ \sqrt{- \hat g} \;  T_b \Bigr]
 \,.
\end{equation}
With the same \ansatz, the corresponding result for the Maxwell
field reduces to
\begin{equation}
 \Bigl[2\pi \sqrt{-\hat g} \; e^{-B + (n-2)W}
 \, f \, A_\theta' \Bigr]_{x_b}
  = \frac{\pd }{\pd A_\theta} \Bigl[ \sqrt{- \hat g}
  \; \, T_b \Bigr]
  := \sqrt{- \hat g} \;J_b(\phi) \,,
\end{equation}
where the last equality defines the quantity $J_b$.

Finally, for fixed-$\rho$ surfaces in this \ansatz, $K_{ij} =
\frac12 \, \partial_\rho g_{ij}$, and the comparison `flat' metric
is $\exd s^2_{\rm flat} = \exd \rho^2 + \rho^2 \exd \theta^2 +
e^{2W_{\rm flat}} \hat g_{\mu\nu} \, \exd x^\mu \exd x^\nu$, with
$W'_{\rm flat} \to 0$ as $\rho \to 0$. Since $K_{\theta\theta} =
B' e^{2B}$ and $K_{\mu\nu} = W' \, e^{2W} \hat g_{\mu\nu}$, we
have $K = g^{ij} K_{ij} = B' + (n-2) W'$, and so the $(\mu\nu)$
components of the metric matching conditions give
\begin{equation} \label{munumatching}
 \left[- \frac{2\pi}{\kappa^2} \sqrt{- \hat g} \, e^{(n-2)W}[ e^{B}
 \left( (n-3) W' +  B'\right) - 1]
 \right]_{x_b} =  \sqrt{-\hat g} \;  T_b(\phi) \,,
\end{equation}
while the $(\theta\theta)$ components are,
\begin{eqnarray} \label{thetathetamatching}
 \left[ \frac{2\pi}{\kappa^2}  \sqrt{-\hat g} \, e^{B + (n-2)W}
 \left( (n-2) \, W' \right) \right]_{x_b} &=& -2
 \frac{\partial}{\partial g_{\theta\theta}} \Bigl[
\sqrt{-\hat g} \, T_b \Bigr] \\
 &:=&  (n-2)\sqrt{-\hat g} \,  U_b(\phi) \,, \nn
\end{eqnarray}
where the last equality defines $U_b$. Just as $T_b$ physically
represents the brane tension, $J_b$ can be interpreted as
describing microscopic axial currents within the brane, or
equivalently any microscopic magnetic flux these currents enclose
within the brane. Once the dimensions transverse to the brane are
dimensionally reduced, $U_b$ turns out \cite{uvcaps,cod2matching}
to be related to the brane contribution to the scalar potential
within the low-energy 4D effective theory defined below the KK
scale (as is seen in more detail later).

\subsection{The brane constraint}

These matching conditions, when combined with the bulk equations
of motion, imply an important constraint relating the quantities
$T_b$, $J_b$ and $U_b$ \cite{santiago,uvcaps,cod2matching}. This
constraint comes from eliminating second derivatives,
$\partial_\rho^{\,2}$, of the fields from the field equations, and
so can be regarded as the `Hamiltonian' constraint on the initial
data when integrating the field equations in the $\rho$ direction.
When written in the form given above, the relevant combination of
Einstein equations is $(n-2)(\mu\nu)+(\theta\theta)-(\rho\rho)$,
which imply
\begin{eqnarray}
\label{constraint}
 &&(n-3)(n-2) \, (W')^2 + 2(n-2) \, W' B' -
 \mathcal G_{\ssA\ssB} \, {\phi^\ssA}'
 {\phi^\ssB}' \nn\\
 && \qquad\qquad\qquad\qquad\qquad
 -\kappa^2 e^{-2B} f \,(A_\theta')^2
 +e^{-2W}\hat R+2\kappa^2V = 0.
\end{eqnarray}

To turn this into a constraint on brane properties, multiply it
through by $e^{2B+2(n-2)W}$ and take the limit $x \to x_b$, using the
above matching conditions to eliminate the derivatives
${\phi^\ssA}'$, $B'$, $W'$ and $A_\theta'$ in favour of the brane
functions $T_b$, $J_b$ and $U_b$. The required matching conditions
are
\begin{eqnarray} \label{matching}
\Bigl[ e^{B} {\phi^\ssA}'\Bigr]_{x_b}
 &=& e^{-(n-2)W}\cG^{\ssA\ssB} \frac{\partial
 \cT_b}{\partial \phi^\ssB} \quad \hbox{with} \quad
 \cT_b := \frac{\kappa^2 T_b}{2\pi}  \nn\\
 \Bigl[\kappa A_\theta' \Bigr]_{x_b} &=&
   e^{-(n-2)W}\frac{\cJ_b}{f}
  \quad \hbox{with} \quad \cJ_b := \frac{\kappa \,
  e^B J_b }{2\pi} \quad \nn\\
 \Bigl[ e^{B} W' \Bigr]_{x_b}
 &=&  e^{-(n-2)W}\cU_b \quad \hbox{with} \quad
 \cU_b :=  \frac{\kappa^2 \, U_b}{2\pi}  \\
 \hbox{and } \quad
 \Bigl[ e^B B'-1 \Bigr]_{x_b}
 &=& -  e^{-(n-2)W}\Bigl[ \cT_b +  (n-3)\,\cU_b
 \Bigr] \,,\nn
\end{eqnarray}
where each of $\cU_b$, $\cT_b$ and $\cJ_b$ is dimensionless
(keeping in mind $e^B$ has dimensions of length). Using
eqs.~\pref{matching} in eq.~\pref{constraint} we find the desired
constraint:
\begin{eqnarray}
\label{curvature constraint1}
&&(n-3) (n-2)\,(\cU_b)^2 + 2(n-2) \cU_b \Bigl[ e^{(n-2)W} -
 \cT_b - (n-3) \, \cU_b \Bigr] \\
 &&\qquad\qquad\qquad\qquad\qquad -
 \mathcal G^{\ssA\ssB} \, \frac{\partial
 \cT_b}{\partial \phi^\ssA} \, \frac{\partial
 \cT_b}{\phi^\ssB} - \frac{(\cJ_b)^2}{f}
 +e^{2B+2(n-2)W} \Bigl[e^{-2W} \hat R+ 2\kappa^2V
 \Bigr]_{x_b} = 0 \,. \nn
\end{eqnarray}

This crucially simplifies once we use the fact that near the brane
$e^B \to 0$ as $\rho \to 0$. (This states that the circumference
of small circles about the brane must vanish as the radius of the
circles vanishes. If not true, the object at $\rho = 0$ would not
be interpreted as a codimension-2 brane.) The key observation
\cite{uvcaps,cod2matching} is that the quantities $\kappa \,
e^{2B} J_b$, $e^{2B-2W} \hat R$ and $\kappa^2 e^{2B} V$ also tend
to vanish in this limit (as would be true, for instance, if
$e^{-2W}\hat R$, $V$ and $J_b$ were bounded at the brane
positions), implying that the constraint becomes
\begin{equation}
\label{curvature constraint}
  (n-2)\cU_b \Bigl[ 2e^{(n-2)W}  - 2\cT_b - (n-3) \, \cU_b \Bigr]
  - (\cT_b')^2 \simeq 0 \,,
\end{equation}
where $(\cT_b')^2 = \mathcal G^{\ssA\ssB} \, \partial_\ssA \cT_b
\, \partial_\ssB \cT_b$.

What is important about this last form of the constraint is that
the on-brane curvature drops out in this limit, meaning that
eq.~\pref{curvature constraint} cannot be read as being solved for
$\hat R$. Instead, this constraint expresses a consistency
condition for the brane action and junction conditions, imposed by
the bulk equations of motion. In practice it provides a very
simple method for computing the quantity $\cU_b(\phi)$ once
expressions for $\cT_b(\phi)$ are given, since solving
eq.~\pref{curvature constraint} implies
\begin{equation} \label{Ubsoln}
 \cU_b = \frac{1}{n-3} \left[ \left( e^{(n-2)W}
 - \cT_b\right) \pm \sqrt{ \left(e^{(n-2)W} - \cT_b\right)^2 -
 \left( \frac{n-3}{n-2} \right) \, (\cT_b')^2 } \right] \,.
\end{equation}
Here the root is chosen for which $\cU_b \to 0$ when $(\cT_b')^2
\to 0$, and so is $\pm$ according to whether sign $\left(
e^{(n-2)W} - \cT_b \right)$ is $\mp$. This means that $\cU_b$ has
the same sign as does $\left( e^{(n-2)W} - \cT_b \right)$. Notice
also that requiring the square root never be complex requires
\be \label{notcomplex}
 \frac{n-3}{n-2} \, \left( \cT_b' \right)^2 \le \left( e^{(n-2) W}
 - \cT_b \right)^2 \,.
\ee
This last condition can be nontrivial, even though control over
the semiclassical approximation requires $|\cT_b| \ll 1$ and
$(\cT_b')^2 \ll 1$. This is because it can happen that $e^W \to 0$
at the brane, in which case eq.~\pref{notcomplex} becomes a
constraint on the size of $(\cT_b')^2/\cT_b^2$.

For $(\cT_b')^2 \ll \left( e^{(n-2)W} - \cT_b \right)^2$
eq.~\pref{Ubsoln} becomes
\begin{equation} \label{Ubsoln1}
 \cU_b  \simeq \frac{(\cT_b')^2}{2(n-2)
 \left(e^{(n-2)W} - \cT_b\right)}
 + \frac{(n-3)(\cT_b')^4}{8(n-2)^2
 \left(e^{(n-2)W} - \cT_b\right)^3} + \cdots \,.
\end{equation}

\subsection{The classical low-energy on-brane effective action}
\label{sec:onbraneaction}

Over distances much longer than the size of the two compact
dimensions transverse to the brane the classical bulk dynamics is
governed by the motion of the massless Kaluza-Klein states. The
dynamics are effectively $d$-dimensional, with $d = n-2$. To
understand the dynamics from this $d$-dimensional perspective, it
is useful to integrate out the extra dimensions to obtain the
low-energy lower-dimensional effective theory. At the classical
level this amounts to eliminating all of the massive KK states as
functions of their massless counterparts, using the bulk classical
equations of motion.

In the present instance the massless KK states consist of the
on-brane metric and Maxwell fields, $\hat g_{\mu\nu}$ and $A_\mu$,
as well as any $d$-dimensional scalars, $\varphi^a$, descending
from $\phi^\ssA$ and/or from moduli in the metric components,
$g_{mn}$, in the extra dimensions. To obtain the low-energy
potential, $V_{\rm \,eff}(\varphi)$, for the various
$d$-dimensional scalars, $\varphi^a$, we eliminate the massive
Kaluza-Klein modes in the action, as functions of $\hat
g_{\mu\nu}$ and $\varphi^a$. The transverse metric, $g_{mn}$, is
eliminated by using the trace reversed $(mn)$ Einstein equations,
which single out the kinetic terms for $g_{mn}$:
\begin{align}
\label{gmneom}
 \frac{1}{2\kappa^2} \left( \mathcal{R}_{mn}
 + \mathcal G_{\ssA \ssB} \, \partial_m \phi^\ssA
 \partial_n \phi^\ssB \right) +
 \frac{f}{2} \,  \, F_m^{\ \ \ssP} F_{n \ssP}
 + \frac{1}{n-2} \left[V - \frac{f}{4} \, F_{\ssP\ssQ}
 F^{\ssP\ssQ} \right] \,  g_{mn} &= 0,
\end{align}
These comprise two independent equations, which we take to be the
sum and difference of the $(\rho\rho)$ and $(\theta\theta)$
components. The difference gives
\be \label{R2eomdiff}
 (n-2)\Bigl( W''+(W')^2 - W' B' \Bigr)
  + \cG_{\ssA\ssB} \, \phi^\ssA{}' \phi^\ssB{}' = 0\,,
\ee
while the sum is equivalent to contracting eq.~\pref{gmneom} with
$g^{mn}$, to give
\be \label{R2eom}
 \frac{1}{2\kappa^2}\left( \, \cR_{(2)} + \mathcal G_{\ssA \ssB}\pd_m\phi^\ssA\pd^m\phi^\ssB\right)
 = -\frac{n-3}{2(n-2)} \, f \, F_{mn} F^{mn} - \frac{2}{n-2}V \,,
\ee
where we write the higher-dimensional curvature scalar as
\bea
 \mathcal R = g^{\ssM \ssN} {\mathcal R^\ssP}_{\ssM
 \ssP \ssN} &=& \cR_{(n-2)} + \cR_{(2)} \nn\\
 \hbox{where} \quad
 \cR_{(2)} = g^{mn} {\cR^\ssP}_{m \ssP\,n} &=& R_{(2)} + (n-2)
 ( \Box W + \nabla W \cdot \nabla W) \nn\\
 &=& R_{(2)} + (n-2)
 \Bigl[W'' + (W')^2 + B' W' \Bigr] \label{HDReqn0}\\
 \hbox{and} \quad
 \cR_{(n-2)} = g^{\mu\nu}
 {\cR^\ssP}_{\mu\ssP\nu} &=& e^{-2W}
 \hat g^{\mu\nu}  \hat R_{\mu\nu}
 + (n-2) [\Box W + (n-4) \nabla W \cdot \nabla W] \nn\\
 &=& e^{-2W} \hat g^{\mu\nu}  \hat R_{\mu\nu}
 + (n-2) \Bigl[W'' + (n-4) (W')^2 + B' W' \Bigr]
 \,.\nn \label{HDReqn}
\eea
Here $R_{(2)} = g^{mn} {R^p}_{mp\,n}$ and $\hat g^{\mu\nu} \hat
R_{\mu\nu}$ respectively denote the curvature scalars built from
the 2D metric, $g_{mn}$, and the 4D metric, $\hat g_{\mu\nu}$.

Using eq.~\pref{R2eom} to eliminate $\cR_{(2)}$ from the bulk
action then yields the bulk contribution to the lower-dimensional
lagrangian density.\footnote{Although in principle the
extra-dimensional part of the trace reversed $(\mu\nu)$ Einstein
equation, $ER_{\mu\nu}(x,y) = 0$ could also be used to eliminate
massive KK modes, this {\em cannot} be used to eliminate
$R_{(n-2)}$ from $V_\ssB$ because the integration in
eq.~\pref{VBexpn} projects onto the zero-mode component of
$E_{\mu\nu} = 0$.} Using $\sqrt{-g} = \sqrt{- \hat g} \;
\sqrt{g_2} \; e^{(n-2)W}$, we find

\begin{eqnarray} \label{VBexpn}
  \cL_{\text{eff}}(\varphi) &=& -\int \exd^2 x \sqrt{g_2} \;
 e^{(n-2)W} \left[ \frac1{2\kappa^2} \, \cR_{(n-2)}
  + \frac{4-n}{4(n-2)}  \, f \, F_{mn} F^{mn}
  +\frac{n-4}{n-2} \, V \right] \nn\\
  &=& -\int \exd^2 x \sqrt{g_2} \;
 e^{(n-2)W} \left\{ \frac1{2\kappa^2} \, \Bigl[
 e^{-2W} \hat g^{\mu\nu}  \hat R_{\mu\nu}
 + (n-2) \Bigl(W'' + (n-4) (W')^2 + B' W' \Bigr)\right.\Bigr] \nn\\
 && \left. \qquad\qquad\qquad\qquad\qquad\qquad\qquad
   + \frac{4-n}{4(n-2)}  \, f \, F_{mn} F^{mn}
   +\frac{n-4}{n-2} \, V \right\}\nn\\
   &=& -\int \exd^2 x \sqrt{g_2} \;
 e^{(n-2)W} \left\{ \frac1{2\kappa^2} \, \Bigl[
 e^{-2W} \hat g^{\mu\nu}  \hat R_{\mu\nu}
 + (n-2) \Bigl( (n-5)(W')^2+2W'B'\Bigr) \right.\nn\\
 && \left. \qquad\qquad\qquad\qquad\qquad\qquad\qquad
 -\mathcal G_{\ssA \ssB}\phi^\ssA{}'\phi^\ssB{}'
 \Bigr]
 + \frac{4-n}{4(n-2)}  \, f \, F_{mn} F^{mn}
 +\frac{n-4}{n-2} \, V \right\}
 \,. \nn\\
 &=&- \int \exd^{n-2} x \sqrt{- \hat g}
 \;\left[\frac1{2\kappa_{\ssN}^2} \, \hat g^{\mu\nu}
 \hat R_{\mu\nu} +V_{B}\right],
\end{eqnarray}
where the second to last equality uses the second independent bulk
field equation, eq.~\pref{R2eomdiff}, the last equality defines
the bulk potential, $V_B$, and the lower-dimensional Newton's
constant, $\kappa^2_\ssN = 8 \pi G_\ssN$, is given by
\be  \label{4d gravitational coupling}
 \frac{1}{\kappa^2_\ssN(\varphi)} := \frac{1}{\kappa^2}
 \int \exd^2 x \sqrt{g_2} \; e^{(n-4)W} \,.
\ee
In general this depends on the low-energy scalar fields, a
dependence that can be removed by performing a Weyl rescaling to
reach the lower-dimension Einstein frame.

To obtain the complete low-energy scalar potential,
$V_{\rm\,eff}$, the bulk contribution, $V_{B}$, must be combined
with two other contributions, both associated with the source
branes. The first of these comes from the boundary terms of the
bulk action \cite{uvcaps,cod2matching}, such as the
Gibbons-Hawking term for the metric, evaluated at a small surface,
$\Sigma_b$, situated a short proper distance, $\rho = \epsilon$,
from the position of each of the source branes:
\begin{eqnarray}
 S_{\GH} &=&
 \sum_{b=0}^1 \lim_{\epsilon \to 0} \oint_{\Sigma_b}
 \exd \theta \, \exd^{n-2} x \, \frac{1}{\kappa^2}
 \sqrt{-\hat\gamma} \; K \nn\\
 &=& \frac{2\pi}{\kappa^2}
 \sum_{b=0}^1 (-)^b \int_{\rho = \rho_b}
 \exd^{n-2} x \sqrt{-\hat g} \; e^{B+(n-2)W} \, \Bigl[
 B' + (n-2) W' \Bigr] \nn\\
  &=& - \sum_{b=0}^1 \int_{\rho = \rho_b}
 \exd^{n-2} x \sqrt{-\hat g}  \, \left\{
 \Bigl[ - T_b - (n-3) U_b \Bigr] + (n-2) U_b \right\}\nn\\
   &=& - \sum_{b=0}^1 \int_{\rho = \rho_b}
 \exd^{n-2} x \sqrt{-\hat g}  \,
 \Bigl( U_b - T_b \Bigr) \,.
\end{eqnarray}
Here we use the axisymmetric {\em ansatz}, as is appropriate very
near the source branes. The relative sign, $(-)^b$, and the
overall sign in the second line arise because primes denote
$\exd/\exd \rho$ while the derivatives appearing in the
Gibbons-Hawking action and matching conditions are outward
directed, and this is in the $\exd \rho$ direction for one brane
and $- \exd \rho$ for the other. The last line uses the matching
conditions described earlier to exchange $W'$ and $B'$ for terms
involving the brane action, using the fact that the contribution
of $[e^B K]_{\rm flat}$ cancels between the two branes.

The second contribution to the 4D scalar potential comes from the
contribution of the brane action itself,
eq.~\pref{cod2braneaction}. Combining these with $V_{\rm 4B}$
above gives the full 4D scalar potential in the classical limit as
in \cite{cod2matching},
\begin{align}
 - \int \exd^{n-2} x \sqrt{-\hat g} \; V_{\rm\,eff}
 =& - \int \exd^{n-2} x \sqrt{- \hat g} \; V_{B}
 +\sum_{b=0}^1 \Bigl[ S_b + \lim_{\epsilon \to 0}
 S_{\rm GH} \Bigr] \\
 =& - \int \exd^{n-2} x \sqrt{- \hat g} \; V_{B}
 - \sum_{b=0}^1 \int \exd^{n-2}x \, \sqrt{-\hat g}
 \left[ T_b +  \Bigr( U_b - T_b
 \Bigr) \right]  \,, \nn
\end{align}
where the notation $W_b$ is a reminder that $W$ is evaluated at
the brane position. This shows that (within the classical
approximation) the effect of the Gibbons-Hawking terms is to
ensure that the net contribution of each brane to the low-energy
scalar potential is given by the quantity $U_b$, appropriately
warped. The complete low-energy scalar potential is therefore,
\begin{eqnarray} \label{Vefffinal}
 V_{\rm\,eff} &=& V_{\ssB} + \sum_b  U_b \nn\\
 &=& \sum_b  U_b + \int \exd^2 x \sqrt{g_2} \;
 e^{(n-2)W} \left\{ \frac1{2\kappa^2} \, \Bigl[
 (n-2) \left\{(n-5)(W')^2+2W'B'
 -\mathcal G_{\ssA \ssB}\phi^\ssA{}'\phi^\ssB{}'\right\}
 \Bigr]\right.\nn\\
 && \left. \qquad\qquad\qquad\qquad\qquad\qquad\qquad
 + \frac{4-n}{4(n-2)}  \, f \, F_{mn} F^{mn}
 +\frac{n-4}{n-2} \, V \right\}
 \,. \nn
\end{eqnarray}

\subsubsection*{Stationary points}

For some purposes it is sufficient to obtain the value of the
potential, $V_{\rm\,eff}(\phi_0)$, evaluated at its stationary
point, where $V_{\rm\,eff}'(\varphi_0) = 0$. This can be obtained
from the higher-dimensional action by eliminating fields using
{\em all} of the equations of motion, and not just those of the
massive KK modes. In this case we may directly use the equation of
motion,
\be
 \frac{1}{2\kappa^2} \, \Bigl( \mathcal R
 + \mathcal G_{\ssA\ssB} \, \partial_\ssM \phi^\ssA
 \, \partial^\ssM \phi^\ssB \Bigr) = - \frac{(n-4)}{
 4(n-2)} \, f \, F_{\ssM\ssN} F^{\ssM\ssN}
 - \frac{n V}{n-2} \,,
\ee
rather than eq.~\pref{R2eom} for $R_{(2)}$. Using this to
eliminate $\mathcal R$ from the bulk action yields
\begin{align} \label{bulk integrated full}
 S_{\rm\,ext} =& -\int \exd^n x \sqrt{-g} \;
 \left[ \frac1{2\kappa^2} \Bigl( \mathcal R
 + \mathcal G_{\ssA \ssB} \, \pd_\ssM \phi^\ssA
 \pd^\ssM \phi^\ssB \Bigr) + \frac14 \, f \,
 F_{\ssM \ssN} F^{\ssM \ssN} + V \right]_{\rm\,cl} \nn\\
 =& - \frac{2}{n-2} \int \exd^n x \sqrt{-g} \;
 \left[
 \frac14 \, f \, F^{mn} F_{mn} -  V  \right] \,.
\end{align}

When comparing with the low-energy theory we must also evaluate
the low energy action at its stationary point. That is, we
evaluate the action
\be
 S_{\rm\,eff} = -\int \exd^{n-2}x \sqrt{-\hat g}
 \; \left[ \frac{1}{2\kappa^2_\ssN}
 \, \hat R_{(n-2)} + V_{\rm\,eff} \right] \,,
\ee
at the solution to the low-energy field equations,
\be
 \frac{1}{2\kappa^2_\ssN} \hat R_{(n-2)}
 = - \frac{(n-2)}{n-4} \, V_{\rm\,eff} \,,
\ee
leading to
\be \label{Sextgeneral}
 S_{\rm\,ext} = \frac{2}{n-4} \int \exd^{n-2}x \,
 \sqrt{- \hat g} \; V_{\rm\,eff}(\varphi_0) \,.
\ee
Using the previous results for $V_{\rm\,ext}$ and the brane
contribution then gives
\be \label{extVresult}
 \frac{2}{n-4} \,V_{\rm\,eff}(\varphi_0) =
 - \sum_b e^{(n-2)W_b} U_b
 - \frac{2}{n-2} \int \exd^2 x \sqrt{g_2} \; e^{(n-2)W}
 \left[ \frac14 \, f \, F^{mn} F_{mn} -  V  \right]\,.
\ee

In many cases of interest the bulk contribution to this expression
can itself also be written as a sum of contributions localized at
the position of each brane. This is true, in particular, whenever
the bulk action, $S_\ssB = \int \exd^n x \; \cL_\ssB$, enjoys a
classical scaling symmetry, under which $\cL_\ssB[\lambda^{p_i}
\psi_i] \equiv \lambda \cL_\ssB[\psi_i]$, for arbitrary real,
constant $\lambda$. (This type of scale symmetry generically holds
for higher-dimensional supergravity theories in particular.) When
this is true the lagrange density satisfies the identity
\begin{eqnarray}
 \cL_\ssB &\equiv& \sum_i p_i \left[ \psi_i
 \frac{\partial \cL_\ssB}{\partial
 \psi_i} + \partial_\mu \psi_i \,
 \frac{\partial \cL_\ssB}{\partial (\partial_\mu \psi_i)}
 \right] \nn\\
 &=& \sum_i \left\{\partial_\mu \left[ p_i \frac{\partial
 \cL_\ssB}{ \partial \partial_\mu \psi_i} \right]
 + p_i \psi_i
 \left[ \frac{\partial \cL_\ssB}{\partial
 \psi_i} -  \partial_\mu \left(
 \frac{\partial \cL_\ssB}{\partial (\partial_\mu \psi_i)}
 \right) \right] \right\} \,,
\end{eqnarray}
which shows \cite{sledxd} that the action becomes a total
derivative whenever it is evaluated at an arbitrary classical
solution. Whenever this is true the entire low-energy potential
can be interpreted as the sum over brane contributions, much as
was done for the Gibbons-Hawking term above.

\section{Examples}

It is instructive to test the above construction by applying it to
situations for which explicit solutions are known for the
higher-dimensional theory. We do so in this section using F-theory
compactifications of 10D Type IIB supergravity to 8 dimensions in
the presence of space-filling D7 branes, and using
compactifications to 4 dimensions of supersymmetric and
nonsupersymmetric six-dimensional theories.

\subsection{D7 branes in F-Theory}

We start with F-theory \cite{ftheory} compactifications of Type
IIB supergravity to 8 dimensions, which serves as an example where
explicit forms for the bulk and brane actions are known, as are
closed-form expressions for the bulk sourced by various
space-filling brane configurations \cite{scs}. This provides a
check on the validity of the matching conditions, and on the
low-energy on-brane scalar potential.

The bulk fields to be followed in this case are the metric,
$g_{\ssM\ssN}$, and the axio-dilaton,
\be
 \tau = C_0 + i \, e^{-\phi} \,,
\ee
where $C_0$ is the Ramond-Ramond scalar and $\phi$ is the 10D
dilaton, for which the string coupling is $g_s = e^\phi$. The bulk
action for these fields in the 10D Einstein frame is
\begin{equation} \label{ftheoryaction}
 S_\ssB = - \frac{1}{2\kappa^2}
 \int \exd^{10}x \sqrt{-g} \; g^{\ssM \ssN}
 \left[ \cR_{\ssM \ssN} + \frac{\partial_\ssM \overline\tau
 \,\partial_\ssN \tau}{2\, (\hbox{Im}\, \tau)^2} \right] \,,
\end{equation}
which is invariant under PSL(2,$R$) transformations
\be
 \tau \to \frac{a \tau + b}{c \tau + d} \,,
\ee
with the real parameters $a$ through $d$ satisfying $a\,d-b\,c =
1$. Quantum effects are expected to break this to PSL(2,$Z$), for
which the parameters are restricted to be integers. Since $e^\phi
\ge 0$ the field $\tau$ lives in the upper-half $\tau$ plane, but
because of the symmetry it suffices to consider $\tau$ to live
within the fundamental domain, $\cF$, defined by modding out the
upper half plane by a PSL(2,$Z$).

\subsubsection*{Bulk solutions}

The scalar field equation for this action is
\be
 \partial \overline\partial \, \tau + \frac{2 \, \partial \, \tau
 \, \overline \partial\, \tau}{\overline \tau - \tau} = 0
 \,,
\ee
which is satisfied by any holomorphic function, $\tau = \tau(z)$,
for which $\overline \partial \,\tau = 0$.

Explicit solutions to the field equations to this model are known
\cite{scs}, for which two of the dimensions are compactified.
Using complex coordinates, $z = x^8 + i x^9$, for the compact
dimensions, the solutions are given by
\be
 j(\tau(z)) = P(z) \quad \hbox{and} \quad
 \exd s^2 = \eta_{\mu\nu} \, \exd x^\mu \exd x^\nu
 + e^{2C(z,\overline z)} \, \exd \overline z \, \exd z \,,
\ee
where the properties of the functions $j(\tau)$, $P(z)$ and
$C(z,\overline z)$ are now described.

The function $j(\tau)$, is the standard bijection from the
fundamental domain, $\cF$, to the complex sphere, given in terms
of Jacobi $\vartheta$-functions by
\be
 j(\tau) = \frac{1728 \, \left[ E_4(\tau) \right]^3}{
 \left[ E_4(\tau) \right]^3 - \left[ E_6(\tau) \right]^2} \,,
\ee
where $E_k(\tau)$ are the Eisenstein modular forms \cite{koblitz}.
For large $\hbox{Im}\, \tau$, $j(\tau)$ diverges zero
exponentially quickly, and the factor of 1728 is chosen so that it
asymptotes to $j(\tau) \simeq e^{-2\pi i \tau} + \cdots$.

$P(z)$ is a holomorphic function, whose singularities occur at the
locations of the source branes, $z = z_i$ for $i = 1,...,N$. Since
the singularities of the metric turn out to be conical when $P(z)$
has isolated poles as $z \to z_i$, it is convenient to choose
$P(z)$ to be a ratio of polynomials. The simplest case could be
taken as $P = 1/z$, describing a source at $z = 0$, but it turns
out that the metric obtained from the Einstein equations is not
compact in this case. The metric is compact when $P(z)$ has 24
zeroes, such as for the choice
\be
 P(z) = \frac{4 (24 f)^3}{27 g^2 + 4 f^3} \,,
\ee
with $f(z)$ a polynomial of degree 8 and $g(z)$ a polynomial of
degree 12. This gives a compactification of Type IIB supergravity
on $CP^1$, corresponding to an F-theory reduction on $K3$ \cite{ftheory}.

Finally, the metric function $C(z,\overline z)$ is chosen by
solving the Einstein equation. Using $\cR_{z\overline z} = 2
\,\partial \overline \partial \, C$ and $\overline \partial \,\tau
= 0$, this equation of motion is
\be \label{holoeinstein}
 2\, \partial \overline \partial \, C = \frac{\partial \, \tau
 \overline \partial \, \overline \tau}{(\tau - \overline \tau)^2}
 = \partial \overline \partial \, \ln \Bigl(\hbox{Im} \, \tau
 \Bigr)\,.
\ee
The required solution is
\be \label{metricftheory}
 e^{2C(z,\overline z)} = (\hbox{Im}\, \tau) \, \left| \eta^2(\tau) \,
 \prod_{i = 1}^N (z - z_i)^{-1/12} \right|^2 \,,
\ee
where $\eta(\tau) = q^{1/24} \prod_k (1 - q^k)$, for $q = e^{2\pi
i \tau}$, denotes the Dedekind $\eta$-function, and the product
runs over the singularities of $P(z)$. The first factor of this
expression is chosen to satisfy eq.~\pref{holoeinstein}, and the
holomorphic factors are chosen to ensure invariance under
PSL(2,$Z$), and by the requirement that the result does not vanish
anywhere.

\subsubsection*{Brane sources}

The presence of branes in these solutions is signaled by
singularities where $P(z) \simeq c_i/(z - z_i)$, for which $q =
e^{2\pi i \tau} \simeq (z-z_i)/c_i$, and so the above solution
implies
\bea
 \tau(z) &\simeq& \frac{1}{2\pi i} \, \ln (z-z_i) + \cdots \nn\\
 \hbox{and} \quad
 e^{2C(z,\overline z)} &\simeq& k \; \hbox{Im} \,
 \tau  \,,
\eea
for constant $k$. As $z \to \infty$, on the other hand, $P(z)$
remains bounded and so $\tau$ approaches some finite value. In
this case the metric function becomes
\be
 e^{2C(z,\overline z)} \propto (z \overline z)^{-N/12} \,,
\ee
and so if we change coordinates to $z = 1/w$ we have $e^{2C} \,
\exd z \, \exd \overline z \simeq |w|^{(N-24)/6} \, \exd w \, \exd
\overline w$, which is nonsingular because $N = 24$. But each
individual brane contributed to this an amount $e^{2C} \simeq
|w|^{1/6} \, \exd w \, \exd \overline w \propto r^{1/6} \, (\exd
r^2 + r^2 \exd \theta^2)$, which we saw below
eq.~\pref{metricansatz} corresponds to a deficit angle of $\delta
= \pi/6$.

\subsubsection*{Matching conditions}

We are now in a situation to use these solutions to test the
matching conditions found in earlier sections. We can do so even
though the geometry involved is not axisymmetric, because it
becomes effectively axisymmetric in the near-brane limit.

To this end we assume a brane action of the form
\be
 S_b = - \int \exd^8 x \sqrt{- \gamma} \; T_b(\tau, \overline
 \tau) \,,
\ee
where for a D7-brane in the Einstein frame we expect
\be
 T_b = T_* \, e^{\phi} = \frac{ T_*}{ \hbox{Im} \,\tau}
 = \frac{2i \, T_*}{\tau - \overline \tau} \,,
\ee
for constant $T_*$.

Keeping in mind that $W = 0$ for the bulk solutions given above,
the matching condition for the bulk scalar,
eq.~\pref{scalarmatching}, becomes
\begin{equation} \label{scalarmatchingtau}
  \frac{2\pi}{\kappa^2}\left[
 \, \frac{e^B}{4 \, (\hbox{Im} \, \tau)^2}
 \;  \partial_\rho \tau  \right]_{x_b}
 =   \frac{2\pi}{\kappa^2}\left[
 \, \frac{r}{4 \, (\hbox{Im} \, \tau)^2}
 \; \partial_{\,r} \tau \right]_{x_b}
 = \frac{\pd \, T_b }{\pd \,\overline \tau}
 = \frac{T_*}{2i \,  (\hbox{Im} \,\tau)^2}\,.
\end{equation}
This uses the change of variables $\exd \rho = e^C \, \exd r$ and
$e^B = r \, e^C$ to convert from proper distance to
conformally-flat coordinates near the brane. Using the near-brane
limit $\tau \simeq \ln r /2\pi i$ to evaluate $[r \, \partial \,
\tau/\partial \, r]_{x_b} \simeq 1/(2\pi i)$, we find the matching
condition becomes $T_* = 1/(2\kappa^2)$.

Notice that since $e^\phi$ is the string coupling constant, this
semiclassical reasoning presupposes $\hbox{Im} \, \tau =
e^{-\phi}$ is large near the brane, so that $\kappa^2 T_b =
\kappa^2 T_*/\hbox{Im}\, \tau = 1/(2\, \hbox{Im}\, \tau) \ll 1$.
This is automatically satisfied as $r \to 0$ because $\hbox{Im}\,
\tau \simeq - (\ln r)/2\pi$.

The metric matching conditions can be understood in a similar way.
First, matching the on-brane components of the metric gives, from
eq.~\pref{munumatching}
\begin{equation} \label{munumatchingtau}
 - \frac{2\pi}{\kappa^2} \Bigl[ e^{B}  \partial_\rho B - 1
 \Bigr]_{x_b} =  - \frac{2\pi}{\kappa^2} \Bigl[r \, \partial_r B - 1
 \Bigr]_{x_b} = - \frac{2\pi}{\kappa^2} \Bigl[r \, \partial_r C
 \Bigr]_{x_b} = T_b(\tau, \overline \tau) = \frac{ T_*}{
 \hbox{Im}\, \tau } \,,
\end{equation}
which again uses $e^B \partial_\rho = r \, \partial_r$ as
well as $B = C + \ln r$. Using eq.~\pref{metricftheory} gives
$e^{2C} \simeq \hbox{Im}\, \tau $ near the brane, and so $r \,
\partial_r C \simeq \frac12 \, (r \, \partial_r \hbox{Im} \,
\tau)/\hbox{Im} \, \tau$ to get $\left[ r \, \partial_r C
\right]_{x_b} = - 1/(4\pi \,\hbox{Im}\, \tau)$. Once again the
dependence on $\hbox{Im}\, \tau$ is consistent on both sides and
so the matching condition boils down to the statement $2 \kappa^2
T_* = 1$, as above.

A further check comes from using the values for $\kappa^2$ and
$T_*$ for a D7-brane predicted in string theory \cite{polchinski}.
Using $T_* = 2\pi/\ell_s^8$ and $\kappa^2 = \ell_s^8/4\pi$, where
$\ell_s = 2\pi \sqrt{\alpha'}$ is the string length, we have
\begin{equation}
 2 \kappa^2 T_* = 2 \left( \frac{\ell_s^8}{4\pi} \right)
 \left( \frac{2\pi}{\ell_s^8} \right) = 1 \,,
\end{equation}
as required.

Finally, the absence of warping in the bulk solution --- $W = 0$
--- implies that the remaining metric matching condition,
eq.~\pref{thetathetamatching}, degenerates to $U_b = 0$. To
compute $U_b$ in the present instance we use the constraint,
eq.~\pref{Ubsoln}, specialized to $n=10$ dimensions
\begin{equation} \label{Ubsolntau}
 \cU_b = \frac17 \left[ (1 -\cT_b) - \sqrt{ (1-\cT_b)^2 -
 \frac78 \, (\cT_b')^2 } \right] \,,
\end{equation}
where $\cT_b = \kappa^2 T_b/2\pi = \kappa^2 T_*/(2\pi \,
\hbox{Im}\, \tau)$, and use
\be
 (\cT_b')^2 = 2 \, (\hbox{Im}\, \tau)^2  \frac{\partial \cT_b}{\partial
 \tau} \, \frac{\partial \cT_b}{\partial \overline \tau}
 = \frac{1}{2 \, (\hbox{Im}\,
 \tau)^2}  \left( \frac{\kappa^2 T_*}{2\pi} \right)^2
 = \frac{1}{8\pi^2 \, (\hbox{Im}\, \tau)^2} \,.
\ee
Clearly $(\cT_b')^2 = 0$ because $\hbox{Im}\, \tau \to \infty$ as
one approaches the brane, and this in turn ensures $U_b = 0$, as
desired.

As a final check we compute the effective scalar potential,
$V_{\rm\,eff}$, for the KK scalar zero mode in the 8D theory on
the brane, after dimensional reduction. Because $U_b = 0$ this
simply amounts to evaluating the action, eq.~\pref{ftheoryaction},
at the classical solution to the extra-dimensional Einstein
equations, which state
\be
 \cR_{mn} + \frac{1}{4 \, (\hbox{Im}\, \tau)^2}
 \Bigl[  \partial_m \, \tau \, \partial_n \, \overline
 \tau + \partial_n \, \tau \, \partial_m \, \overline
 \tau \Bigr] = 0 \,.
\ee
We see that $V_{\rm\,eff} = 0$ in the effective theory, which is
consistent with the maximally symmetric on-brane geometry being
flat.

\subsection{Brane-axion couplings in 6D}
 \label{sec-rugbyexample}

We next apply the above matching conditions to the example of two
branes coupled to a bulk Goldstone mode (axion), $\phi$, in six
dimensions. Since 6D examples with flat on-brane geometries are
already discussed in some detail in refs.~\cite{cod2matching}, we
concentrate here on solutions to the higher-dimensional equations
for which the on-brane geometry is known to be curved. Our
purposes is to provide a nontrivial example for which the shape of
the full low-energy potential, $V_{\rm\,eff}(\varphi)$, and its
value at its stationary point, $V_{\rm\,eff}(\varphi_0)$, can be
computed explicitly directly from the higher-dimensional theory.
Because this allows a check on how $V_{\rm\,eff}$ varies from its
minimum, it allows us to verify that the extremal point is
actually a local minimum of the low-energy potential.

The simplest such a system starts with gravity coupled to a single
bulk scalar and Maxwell field, with the bulk lagrangian density
given by,
\begin{equation}
 \cL_B = -  \sqrt{-g} \;
 \left\{ \frac{1}{2\kappa^2} \, g^{\ssM\ssN}\Bigl[
 \mathcal{R}_{\ssM\ssN} + \partial_\ssM \phi
 \, \partial_\ssN \phi \Bigr] + \frac14
 \, F_{\ssM \ssN} F^{\ssM \ssN} + \Lambda \right\} \,,
\end{equation}
where $\Lambda$ is a bulk cosmological constant whose value can be
chosen to obtain any desired curvature on the brane. Notice that
the choices $f(\phi) = 1$ and $V(\phi) = \Lambda$ ensure the
action has a shift symmetry, $\phi \to \phi + \xi$, that
guarantees the existence of a scalar KK zero mode having a
constant profile across the bulk. This is the only such
classically massless scalar KK mode, because the presence of the
bulk cosmological term, $\Lambda$, breaks the rigid scaling
symmetry that the Einstein action normally has. This breaking
ensures that the presence of $\Lambda$ removes the `breathing'
mode corresponding to rigid expansions of the extra dimensional
geometry, that would have otherwise have been a low-energy scalar
zero mode.

\subsubsection*{Bulk solutions}

The field equations in this case admit explicit solutions for
which the 4D on-brane geometry is maximally symmetric and the
extra dimensions are axially symmetric \cite{conical,6dnonsusy}.
Using the {\em ansatz} of eq.~\pref{metricansatz}, a simple
solution is
\begin{align}
 \exd s^2 &= \hat g_{\mu\nu} \, \exd x^\mu \exd x^\nu
 + \exd \rho^2 + \alpha^2 L^2 \sin^2\left(
 \frac{\rho}{L}\right) \exd \theta^2 \\
 F_{\rho\theta} &= \alpha \cB_0 L \sin\left(
 \frac{\rho}{L}\right),
\end{align}
with $\phi = \phi_0$ constant. The bulk field equations imply the
following relation amongst the constants $\cB_0$, $L$ and
$\Lambda$:
\begin{equation}
 \label{rugby constraints}
 \cR_{(2)} = - \frac{2}{L^2} = -\kappa^2 \left(
  \frac{3\cB_0^2}{2} + \Lambda \right) \,,
\end{equation}
and the curvature of the on-brane metric is given by
\begin{equation} \label{4Dcurvature}
 {\hat R} = {2\kappa^2} \left( \frac{\cB_0^2}{2}-\Lambda \right).
\end{equation}

When $\alpha = 1$ the extra-dimensional metric describes a sphere
of radius $L$. When $\alpha \ne 1$ the geometry would still look
like a sphere if we redefine $\theta \to \alpha \vartheta$,
although $\vartheta$ is then not periodic with period $2\pi$. This
indicates there are conical singularities at both $\rho = 0$ and
$\rho = \pi L $, with defect angle given by $\delta =
2\pi(1-\alpha)$.

\subsubsection*{Brane properties}

We now ask for a pair of brane sources located at these two
singularities that can support this geometry. We again take
codimension-2 brane actions of the form
\be
 S_b = - \int \exd^4x \sqrt{-\gamma} \; T_b(\phi) \,.
\ee

Because the bulk solution has constant scalar, $\phi = \phi_0$,
its derivative, $\partial_\rho \phi$, vanishes at both branes.
This is only consistent with the scalar matching condition if
$T_b'(\phi)$ also vanishes for both branes when evaluated at the
same place: $\phi = \phi_0$. The vanishing of $T_b'(\phi)$ at
$\phi = \phi_0$ also ensures $U_b(\phi)$ vanishes there, and this
is consistent with the ($\theta \theta$) matching condition,
eq.~\pref{thetathetamatching}, because $W = 0$ throughout the bulk
in the classical solution ensures $\partial_\rho W = 0$ at the
brane positions.

Finally, the ($\mu\nu$) matching condition,
eq.~\pref{munumatching}, reads
\be
 - \frac{2\pi}{\kappa^2} \Bigl[ e^B \, B' - 1 \Bigr]_{x_b} =
 T_b(\phi_0) \,.
\ee
Using $e^B = \alpha L \sin(\rho/L)$ gives $e^B \, B' \to \alpha$
as $\rho \to 0$, and so this matching condition gives the usual
expression for the defect angle in terms of the brane tension,
\be
 \delta = 2\pi(1 - \alpha) = \kappa^2 \,T_b(\phi_0) \,,
\ee
and so $\cT_b = \kappa^2 T_b/2\pi = 1 - \alpha$.

\subsubsection*{The 4D perspective}

We now show how the above picture is reproduced in the low-energy
4D effective theory below the Kaluza-Klein scale. Although we
cannot ask in the low-energy theory about the profiles of bulk
fields within the extra dimensions, we can use it to understand
the curvature, $\hat R$, of the 4D on-brane geometry and the
value, $\phi_0$, to which the low-energy scalar field is fixed.

To this end we explore the scalar potential, $V_{\rm\,eff}$, for
the KK zero mode of the scalar, $\phi$, as it is moved away from
$\phi_0$. To do so requires more information about the shape of
$T_b(\phi)$, so we choose for simplicity,
\be
 T_b(\phi) = M_b^4 + \frac{\mu^4_b}{2} \, (\phi - \phi_0)^2 \,,
\ee
although any choice for $T_b(\phi)$ would do, so long as both
tensions share a common zero for $\partial \, T_b/\partial\,
\phi$.

With this choice we have
\be
 \cT_b = \frac{\kappa^2 M^4_b}{2\pi}
 + \frac{\kappa^2 \mu_b^4}{4\pi} \, (\phi - \phi_0)^2 \,,
 \qquad
 \cT_b{\,'} = \frac{\kappa^2 \mu_b^4}{2\pi} \, (\phi - \phi_0) \,,
\ee
and so to lowest nontrivial order in $\kappa^2$
\bea
 \cU_{\,b} &=& \frac1{3} \left[ (1 - \cT_b) - \sqrt{ (1 - \cT_b)^2 -
 \frac{3}{4} \, (\cT_b{\,'})^2 } \right] \nn\\
 &\simeq& \frac{(\cT_b{\,'})^2}{8(1 - \cT_b)}
 + \frac{3(\cT_b{\,'})^4}{128(1 - \cT_b)^3} + \cdots \,.
\eea
Specialized to the above tension this becomes
\be
 U_b \simeq \frac{\kappa^2 \mu_b^8}{16\pi}
  \, (\phi - \phi_0)^2 + \cdots \,.
\ee
Notice \cite{natbrane} that because $\cU_b$ is quadratic in
$\cT_b{}'$, both it and its derivative $\cU_b'$ naturally vanish
at zeroes of $\cT_b{}'$. Furthermore, the coefficient of $(\phi -
\phi_0)^2$ in $U_{\,b}$ is suppressed relative to the same term in
$T_b$ by an additional power of the small dimensionless factor
$\kappa^2 \mu_b^4/8\pi \ll 1$. The full expression for the
effective potential \pref{Vefffinal} in this case reduces to
\begin{eqnarray} \label{Vefffinalex}
 V_{\rm\,eff} &=& \sum_b  U_b + V_{\ssB}(\phi_0) + \frac12
 \, V_\ssB''(\phi_0) (\phi - \phi_0)^2 + \cdots \nn\\
 &=& \sum_b  U_b + \int \exd^2 x \sqrt{g_2} \;
 e^{4W} \left\{ \, -\frac{1}{8} \, F_{mn} F^{mn}
 +\frac{1}{2}\Lambda \right\}  + \frac12
 \, V_{\ssB}''(\phi_0) (\phi - \phi_0)^2 + \cdots \nn\\
 &=&\sum_b  U_b + \frac{\pi}{2} \left(
 \Lambda-\frac{\cB_0^2}{2}\right) \int_0^{\pi L}d\rho e^B +
 \frac12 \, V_{\ssB}''(\phi_0) (\phi - \phi_0)^2 + \cdots \nn\\
 &=& \left(\Lambda-\frac{\cB_0^2}{2}\right) 2\pi\alpha L^2
  + \frac12 \left[ \, V_{\ssB}''(\phi_0) +
  \sum_b \frac{\kappa^2 \mu_b^8}{8\pi} \right]
  (\phi - \phi_0)^2 + \cdots \,. \nn
\end{eqnarray}
using that both $W'$ and $\phi'$ vanish when $\phi = \phi_0$. More
explicit progress requires the calculation of $V_\ssB''(\phi_0)$,
although this can be expected to be non-negative due if the bulk
solution is stable. This shows that $V_{\rm\,eff}(\phi)$ is
minimized at $\phi= \phi_0$, and this is how the 4D theory
understands the value at which $\phi$ is stabilized.

The value of the potential at this minimum has a direct physical
interpretation, since it sets the value of the 4D curvature
through the 4D Einstein equations. These read, as usual
\be
 \hat R_{\mu\nu} - \frac12 \, \hat R \; \hat g_{\mu\nu}
 - \kappa^2_\ssN V_{\rm\,eff} \, \hat g_{\mu\nu} = 0 \,,
\ee
where the 4D Newton coupling is
\be
 \frac{1}{\kappa_\ssN^2} = \frac{2\pi}{\kappa^2} \int_0^{\pi L}
 \exd \rho \; e^B = \frac{4\pi \alpha L^2}{\kappa^2} \,,
\ee
and so
\be
 \hat R = -4 \kappa^2_\ssN V_{\rm\,eff}(\phi_0) = 2 \kappa^2
 \left(\frac{\cB_0^2}{2} -  \Lambda\right) \,,
\ee
in agreement with the higher-dimensional result,
eq.~\pref{4Dcurvature}. Notice that this agreement requires, in
particular, that the brane tensions $T_b(\phi_0) = M_b^4$ drop out
of the low-energy potential.

Finally, notice that evaluating the potential,
eq.~\pref{Vefffinalex}, at its minimum by evaluating the action at
the classical solution gives a result that agrees with the general
expression \pref{extVresult}, which in the present instance
evaluates to
\bea \label{extVresultnew}
 V_{\rm\,eff}(\varphi_0) &=&
 - \sum_b e^{4W_b} U_b
 - \frac12 \int \exd^2 x \sqrt{g_2} \; e^{4W}
 \left[ \frac14 \, f \, F^{mn} F_{mn} -  V  \right]\nn\\
 &=& \frac12 \left( 4 \pi \alpha L^2 \right)
 \left( \Lambda - \frac{\cB_0^2}{2} \right)\,.
\eea

\subsection{Warped and unwarped supersymmetric examples}

A large class of examples of explicit flux compactifications with
nontrivial warping and scalar profiles in the extra dimensions is
provided by solutions
\cite{sled,sledxd,GGP,GGPplus,6DdSSG,6dsusy,scalingpaper} to
chiral 6D supergravity \cite{NS}. Our goal with this example is to
identify the properties of the branes that are required to source
the known solutions. In general the existence of solutions hinges
on the consistency of these brane properties with the form of the
intervening bulk, but these solutions are not known in closed form
in the case where the on-brane dimensions are curved. In this
situation it is much easier to investigate the existence of
solutions using the equivalent formulation in terms of minima of
the low-energy scalar potential, since it is much easier to
determine when such solutions exist.

The solutions of interest take as their starting point the
following bosonic part of the supersymmetric action
\begin{equation}
 \cL_B = -  \sqrt{-g} \;
 \left\{ \frac{1}{2\kappa^2} \, g^{\ssM\ssN}\Bigl[
 \mathcal{R}_{\ssM\ssN} + \partial_\ssM \phi
 \, \partial_\ssN \phi \Bigr] + \frac14 \, e^{-\phi}
 \, F_{\ssM \ssN} F^{\ssM \ssN} + \frac{2g^2}{\kappa^4} \;
 e^\phi \right\} \,,
\end{equation}
where the constant $g$ denotes the 6D gauge coupling for the
Maxwell field. Because this lagrangian enjoys the property
$\cL_\ssB \to \lambda^2 \cL_\ssB$ when $e^{\phi} \to \lambda^{-1}
e^\phi$ and $g_{\ssM \ssN} \to \lambda g_{\ssM \ssN}$, the
arguments of section \ref{sec:onbraneaction} imply it becomes a
total derivative once evaluated at an arbitrary classical solution
\cite{sledxd}:
\be \label{classicalBoxphi}
 \cL_\ssB(g_{\ssM\ssN}^c, A_\ssM^c, \phi^c) =
 \frac{1}{2\kappa^2} \, \sqrt{-g^c} \; \Box \, \phi^c \,.
\ee

\subsubsection*{Bulk solutions}

For this system it is useful to choose a slightly different metric
{\em ansatz} \cite{GGP},
\be \label{sugraansatz}
 \exd s^2 = \cW^2 \, \hat g_{\mu\nu} \, \exd x^\mu \exd x^\nu +
 a^2 \Bigl( \cW^8 \exd \eta^2 + \exd \theta^2 \Bigr) \,,
\ee
where $a = a(\eta)$, $\cW = \cW(\eta)$ and $\hat g_{\mu\nu}$ is, a
maximally symmetric 4D de Sitter metric, with $\hat R = - 12 H^2$.
With these choices the proper circumference of a circle along
which $\theta$ varies from zero to $2\pi$ at fixed $\eta$ is $2
\pi a(\eta)$, and $\exd \rho = a \cW^4 \exd \eta$. The dilaton is
similarly taken to depend only on $\eta$, $\phi = \phi(\eta)$, and
the Maxwell field is given by $A_\theta = A_\theta(\eta)$, so that
\be
 F_{\eta\theta} = Q \, a^2 \, e^{\phi} \,.
\ee

In this case the content of Maxwell's equations is that $Q$ must
be a constant, while the dilaton and the trace-reversed Einstein
equations become
%
\be
 \phi'' =  \frac{2g^2}{\kappa^2} \, a^2 \cW^8 e^\phi
 -\frac{\kappa^2 Q^2}{2} \, a^2 e^\phi\,,
\ee
and
\bea \label{einsteineqn1}
    &&(\mu\nu): \quad
    \frac{\cW''}{\cW} - \frac{(\cW')^2}{\cW^2}
    + \frac12  \, \phi''
    = \left( \frac{\cW'}{\cW} + \frac12 \, \phi'
    \right)'
    = 3 \, H^2a^2 \cW^6   \\
    \label{einsteineqn2}
    &&(\theta\theta): \quad
    \frac{a''}{a} - \frac{(a')^2}{a^2}
    + \frac12 \, \phi'' = \left( \frac{a'}{a} +
    \frac12 \, \phi' \right)'
    = - \kappa^2 Q^2 \, a^2 e^\phi \,.
\eea
In all of these equations primes denote $\exd/\exd \eta$. The
`Hamiltonian constraint' --- {\em i.e.} the ($\eta \eta$) Einstein
equation --- in these variables is similarly
\be \label{einsteinconstraint}
     \frac12 \, (\phi')^2  - \frac{4\,a'
    \cW'}{a\cW} - \frac{6(\cW')^2}{\cW^2}
    =  \frac{2 {g}^2 }{\kappa^2} \, a^2 \cW^8 e^\phi
    - 6 \, H^2 a^2 \cW^6 -
    \frac{\kappa^2}{2} Q^2 \, a^2 e^\phi  \,.
\ee

The scale invariance of the full 6D field equations under $e^\phi
\to e^\phi/\lambda$ and $g_{\ssM\ssN} \to \lambda g_{\ssM \ssN}$
can be seen from the invariance of the above equations under
\be
 \Bigl\{\phi , a, \cW, H \Bigr\} \to \Bigl\{ \phi + \phi_0 ,
 a\, e^{- \phi_0/2} , \cW, H \, e^{\phi_0/2}  \Bigr\} \,,
\ee
for $\phi_0$ an arbitrary real constant. In the case $H = 0$ this
symmetry implies the existence of a one-parameter family of
classical solutions, and a corresponding flat direction (labelled
by $\phi_0$) that represents a classically massless KK zero mode
coming from a combination of the metric and $\phi$ fields.

The above field equations are written so that their
right-hand-sides tend to zero in the near-brane regions, for which
$a \to 0$. For regions where these right-hand-sides are negligible
the equations simplify to
\be
 \phi'' \simeq \left( \frac{\cW'}{\cW} \right)'
 \simeq \left( \frac{a'}{a} \right)' \simeq 0 \,,
\ee
and so, letting $b=\{0,1\}$ for the branes at
$\eta=\{-\infty,+\infty\}$ respectively,
\be \label{asymptoticpowers}
 \phi \simeq (-)^b q_b \eta \,, \quad
 \cW \simeq \cW_b \, e^{(-)^b \omega_b \eta}
 \quad \hbox{and} \quad
 a \simeq a_b \, e^{(-)^b \alpha_b \eta} \,,
\ee
with different choices for the constants $\alpha_b$,
$\omega_b$ and $q_b$ applying for the two limits, $\eta \to
\pm \infty$. For both asymptotic regions these are related by the
constraint, eq.~\pref{einsteinconstraint}, so that
\be \label{qconstraint}
 q_b^2 = 4\omega_b (2 \alpha_b + 3 \omega_b) \,.
\ee

Notice that it is only consistent in the near-brane limit to
ignore the quantities $a^2 \cW^6$, $a^2 e^\phi$ and $a^2 \cW^8
e^\phi$ on the right-hand sides of eqs.~\pref{einsteineqn1}
through \pref{einsteinconstraint} if
\be \label{inequalities}
 2\alpha_b + 6 \omega_b > 0 \,, \quad
 2\alpha_b + q_b > 0 \quad \hbox{and} \quad
 2\alpha_b + 8 \omega_b + q_b > 0 \,.
\ee
The first of these also guarantees the convergence of the 4D
gravitational constant, which is given by ({\em c.f.} eq.~\pref{4d
gravitational coupling})
\be \label{planckwarp}
 \frac{1}{\kappa^2_\ssN} = \frac{2\pi}{\kappa^2}
 \int_{-\infty}^\infty
 \exd \eta \; a^2 \cW^6 \,.
\ee

Furthermore, since our interest is in solutions where $a \to 0$ at
the positions of the brane sources, we demand $\alpha_b
> 0$. This ensures that the circumference of small circles
encircling the branes vanishes in the limit that the branes are
approached. But if $\alpha_b > 0$, then $\omega_b$ must also be
non-negative. To see this, suppose $\omega_b$ were negative. Then
eq.~\pref{qconstraint} would imply $-2\alpha_b - 3\omega_b > 0$,
and so adding this to the first of eqs.~\pref{inequalities} would
give $\omega_b > 0$, in contradiction with the assumption that it
is negative. By contrast, the constant $q_b$ can take either sign.

Solutions to these equations are known to exist for nonzero $H$
\cite{6DdSSG}, although not yet in an explicit closed form.
Closed-form solutions are known, however, in the special case
where $H$ vanishes, given by \cite{GGP,sledxd}
\bea \label{GGP}
    e^\phi &=& \cW^{-2} e^{\phi_0-\lambda_3 \eta} \nonumber\\
    \cW^4 &=& \left( \frac{\kappa^2 Q\lambda_2}{2g\lambda_1} \right)
    \frac{\cosh[ \lambda_1(\eta - \eta_1)]}{\cosh[ \lambda_2
    (\eta - \eta_2)]} \\
    \hbox{and} \qquad
    a^{-4} &=& \left( \frac{2g\kappa^2 Q^3}{\lambda_1^3 \lambda_2}
    \right) e^{2(\phi_0-\lambda_3\eta)}
    \cosh^3[\lambda_1(\eta - \eta_1)]
    \cosh[ \lambda_2(\eta - \eta_2)]  \,. \nonumber
\eea
Here $\eta_i$ and $\lambda_j$ are integration constants, and there
is no loss of generality in choosing, say, $\lambda_2 \ge 0$. The
equations of motion require the constants to satisfy $\lambda_2^2
= \lambda_1^2 + \lambda_3^2$ --- and so, in particular, $\lambda_2
\ge |\lambda_1|$ (with equality if and only if $\lambda_3 = 0$).
$\phi_0$ is an arbitrary constant corresponding to the scale
invariance associated with the flat direction.

Because the terms involving $H$ in the equations of motion become
negligible in the near-brane limit, the $H=0$ solutions also
provide a more detailed picture of the asymptotic regions at $\eta
\to \pm \infty$. The corresponding metric singularities are
generically curvature singularities, except when $\lambda_3 = 0$,
in which case they turn out to be conical \cite{GGPplus}. The
$\lambda_3 = 0$ solutions include the unwarped, constant-dilaton
`rugby ball' configurations of ref.~\cite{sled} as the special
case where $\eta_1 = \eta_2$. Notice also that the limiting
behaviour is as given in eq.~\pref{asymptoticpowers}, with
\be
 \alpha_b = \frac14 \left[3\lambda_1 + \lambda_2 + 2(-)^b
 \lambda_3 \right] \ge 0\,, \quad
 \omega_b = \frac14 \left( \lambda_2 - \lambda_1 \right)
 \ge 0\,,
\ee
and
\be
 q_b = (-)^{b+1}\lambda_3 - \frac12 \left( \lambda_2
 - \lambda_1 \right) \,.
\ee
Notice that the condition $\omega_b \ge 0$ follows from $\lambda_2
\ge |\lambda_1|$, while $\alpha_b \ge 0$ is a consequence of
\be
 3(\lambda_2 + \lambda_1) - 2 \lambda_3 = \sqrt{\lambda_2 +
 \lambda_1}\left( 3 \sqrt{\lambda_2 + \lambda_1} - 2 \sqrt{
 \lambda_2 - \lambda_1} \right) \ge 0 \,.
\ee

A special role is played by the combination
\be
 \omega_b + \frac{q_b}{2} = (-)^{b+1} \frac{\lambda_3}{2}\,,
\ee
since this dictates the size of the Hubble constant, $H$. This can
be seen by integrating eq.~\pref{einsteineqn1}, and using
eq.~\pref{planckwarp} to obtain \cite{6DdSSG},
\be \label{IntegratedCBA-GGP}
    3 H^2 \int_{-\infty}^\infty \exd\eta \; a^2
    \, \cW^{6} = \frac{3 \kappa^2 H^2}{2\pi\kappa^2_4} =
    \left[ \left(\,\ln \cW + \frac{\phi}{2} \right)'\,
    \right]_{\eta = -\infty}^{\eta = + \infty}
    = - \sum_b \left(\frac{q_b}{2} + \omega_b  \right) \,.
\ee
When evaluated for the solutions of eq.~\pref{GGP}, this reduces
to the Friedmann equation
\be
 H^2 = - \frac{2\pi\kappa^2_4}{3 \kappa^2}
    \sum_b\left[\frac{q_b}{2} + \omega_b  \right]
    = \frac{\kappa^2_4}{3} \left[
    \frac{2\pi}{\kappa^2} \sum_b (-)^b \frac{\lambda_3}{2}
    \right]  = 0 \,
\ee
as required. For more general solutions eqs.~\pref{GGP} hold only
approximately in the near-brane region, so the constant
$\lambda_3$ could differ for the asymptotic region near each
brane.

Notice, in particular, that eq.~\pref{IntegratedCBA-GGP} shows
that $H^2 > 0$ (4D de Sitter space) requires at least one of the
$q_b$ to be negative. Furthermore, choosing $q_b < 0$ is
sufficient to ensure that the contribution to $H^2$ of the
corresponding brane is positive, because
\begin{equation}
 - \left( \frac{q_b}{2} + \omega_b \right)
 = \frac{|q_b|}2 - \omega_b = \sqrt{3\omega_b^2+2\alpha_b\omega_b}
 - \omega_b = \omega_b \left( \sqrt{3+\frac{\alpha_b}{\omega_b}}
 - 1 \right) \ge 0 \,.
\end{equation}
This uses both eq.~\pref{qconstraint} and the property that
$\alpha_b$ and $\omega_b$ are both non-negative.

\subsubsection*{Brane properties}

As usual, the matching conditions relate the asymptotic bulk
solutions to the properties of the source branes. Using $\cW =
e^W$, $a = e^B$ and $a \cW^4 \exd \eta = \exd \rho$, and taking
the brane action to be $S_b = - \int \exd^4 x \sqrt{-\gamma} \;
L_b = - \int \exd^4 x \sqrt{-\hat g} \; T_b$, the scalar matching
condition, eq.~\pref{scalarmatching}, becomes
\begin{equation} \label{scalarmatchingNS}
 \frac{2\pi}{\kappa^2} \, \Bigl[  e^{B + 4W}
 \, \partial_\rho \, \phi \Bigr]_{x_b}
  = \frac{\pd }{\pd\phi} \Bigl[\; e^{4W} \, L_b \Bigr]
 \quad \Longrightarrow \quad \Bigl[ (-)^b
  \partial_\eta \, \phi \Bigr]_{x_b}
  = q_b
 = \frac{\kappa^2}{2\pi} \, \left(
  \frac{\pd \, T_b}{\pd\phi} \right) \,,
\end{equation}
where the sign arises because the direction away from the brane is
$(-)^b \exd \eta$ in the two asymptotic regions. The $( \theta
\theta)$ metric matching condition, eq.~\pref{thetathetamatching},
similarly becomes
\be \label{thetathetamatchingW}
 \frac{2\pi}{\kappa^2} \left[ e^{B + 4W}
 \, \partial_\rho \, W  \right]_{x_b} =
 U_b(\phi)
 \quad \Longrightarrow \quad \left[ (-)^b
  \left( \frac{\partial_\eta \, \cW}{\cW}
 \right) \right]_{x_b} = \omega_b
 = \frac{\kappa^2 U_b}{2\pi} \,.
\ee
Finally, the $(\mu\nu)$ components of the metric matching
conditions are
\begin{equation} \label{munumatchingW}
 - \frac{2\pi}{\kappa^2} \Bigl[ e^{4W}[ e^{B}
 \left( 3 \partial_\rho W + \partial_\rho B\right) - 1]
 \Bigr]_{x_b} = T_b(\phi) \,,
\end{equation}
and so
\begin{equation} \label{munumatchingW}
 \left\{ (-)^b \left[ 3 \left( \frac{ \partial_\eta \cW}{\cW} \right)
 + \left( \frac{ \partial_\eta a}{a} \right) \right] - \cW^4
 \right\}_{x_b} = 3 \omega_b + \alpha_b - \cW^4({x_b})
 = - \frac{\kappa^2 T_b}{2\pi}  \,.
\end{equation}

There are now two qualitatively different cases that are worth
considering separately, depending on whether or not $\omega_b = 0$
or $\omega_b > 0$.

\bigskip\noindent{\em Solutions with only conical singularities:}

\medskip\noindent
If $\omega_b = 0$, then eq.~\pref{qconstraint} implies $q_b = 0$
as well, and so both $\phi$ and $\cW$ asymptote to constants near
the brane. Because $\omega_b = 0$ implies $\cW \simeq \cW_b$ is
constant in the near-brane regime, the behaviour $a \sim
e^{\alpha_b \eta}$ implies the extra-dimensional metric is
proportional to
\be
 e^{2\alpha_b \eta}(\cW_b^8 \exd \eta^2 + \exd \theta^2 )
 = \exd \rho^2
 + \left( \frac{\alpha_b \rho}{\cW_b^4} \right)^2 \exd \theta^2 \,,
\ee
showing that it has only a conical singularity at the brane
position, with defect angle $\delta_b = 2\pi(1-\alpha_b/\cW_b^4)$.

When $\omega_b = q_b = 0$, the matching conditions boil down to
\be \label{omegazeromatching}
 \frac{\kappa^2 T_b'}{2\pi} = \frac{\kappa^2 U_b}{2\pi} = 0
 \quad \hbox{and} \quad
 \delta_b = \frac{\kappa^2 T_b}{\cW_b^4} = \kappa^2 L_b  \,.
\ee
The last of these relates the tension to the size of the conical
defect angle in the usual way, while the first states that the
value taken by $\phi$ near each brane must be at a stationary
point of the tension on that brane. (Since this is also
automatically a zero of $U_b$, the second condition is redundant.)
In order for solutions to exist the two tensions must be related
to one another by the known asymptotic limits of the given bulk
solution. That is, if $\phi_b = \lim \phi(\eta)$ as $\eta\to
-(-)^b \infty$, then $T_b$ must satisfy $T_b'(\phi_b) = 0$ at both
ends.

Since its right-hand-side is non-negative, eq.~\pref{einsteineqn1}
shows that it is only possible to have $\omega_b = q_b = 0$ at
{\em both} branes if $H = 0$. If $H = 0$ the solutions given in
eqs.~\pref{GGP} have this property (for both branes) when
$\lambda_3 = 0$ (and so also $\lambda_1 = \lambda_2 := \lambda$).
Notice that $\cW$ and $e^\phi = \cW^{-2}$ need not be identically
constant in this case unless $\eta_1 = \eta_2$.

From the point of view of the 4D theory the result $H = 0$ is
understood for these solutions in terms of the vanishing of the
classical low-energy 4D effective potential,
\be
 V_{\rm\,eff} = V_\ssB + \sum_b  U_b = 0\,.
\ee
This vanishes because eq.~\pref{classicalBoxphi} (when $\phi' =
0$ near the branes) shows that the bulk contribution to the low
energy potential vanishes, $V_\ssB = 0$, and
eq.~\pref{omegazeromatching} implies $U_b = 0$ for both branes.

If $T_b'$ should vanish identically, then so must also $U_b$ and
$V_{\rm\, eff}$. In this case the vanishing of $V_{\rm\,eff}$
shows that the flat direction, corresponding to the scaling $\phi
\to \phi + \phi_0$ and $g_{\ssM\ssN} \to e^{-\phi_0}
g_{\ssM\ssN}$, is not lifted by the classical couplings to the
branes. But if $T_b$ depends nontrivially on $\phi$, then $U_b$
becomes nonzero as soon as $\phi$ differs from its asymptotic
value $\phi_b$, implying that $V_{\rm\,eff}$ depends nontrivially
on $\phi_0$. Since $U_b(\phi_0)$ is given by
\be
 \cU_b = \frac13 \left[ (\cW^4 - \cT_b) - \sqrt{(\cW^4 - \cT_b)^2
 - \frac34 \, (\cT_b{\,'})^2} \right]  \,,
\ee
where $\cT_b = \cT_b(\phi_b + \phi_0)$, it is non-negative
(provided $\cT_b < \cW^4$). Because the bulk action is known to be
stable against small fluctuations about the bulk solutions
\cite{stability}, it follows that $V_{\rm\,eff}(\phi_0)$ must be
minimized by any configuration for which it vanishes, such as
$\phi_0 = 0$ (which corresponds to $\lim \phi = \phi_b$). This
shows how the 4D theory sees that the flat direction, $\phi_0$, of
the bulk equations becomes fixed at the same value as is chosen by
the matching conditions when viewed from the higher-dimensional
perspective.

\bigskip\noindent{\em Solutions with $\omega_b > 0$}

\medskip\noindent On the other hand, if $\omega_b > 0$ then $e^W =
\cW \to 0$ as the brane is approached. In this case the scalar and
$(\mu\nu)$ matching conditions are
\be
 q_b = \frac{\kappa^2 T_b'}{2\pi} = \cT_b{\,'}
 \quad \hbox{and} \quad
 3 \, \omega_b + \alpha_b
 = - \frac{\kappa^2 T_b}{2\pi} = - \cT_b \,.
\end{equation}
Since $\alpha_b$ and $\omega_b$ are both positive, the last of
these conditions implies $T_b < 0$. The third matching condition
in this case is
\be
 \omega_b = \frac{\kappa^2 U_b}{2\pi} = \cU_b
 = \frac13 \left[ -\cT_b - \sqrt{\cT_b^2 - \frac34 \,
 (\cT_b{\,'})^2} \right] \,,
\ee
which also requires $\cT_b < 0$ if $\cU_b$ and $\omega_b$ are to
be positive.

Because we use coordinates for which the branes are situated at
$\eta \to \pm \infty$, we demand that these matching conditions be
satisfied as identities in $\eta$ in the asymptotic regimes. Use
of the asymptotic forms for the bulk solutions in this regime
corresponds to expanding the brane tension about the value taken
by $\phi$ at the brane.

This determines the functional form for the brane action,
$T_b(\phi,a,W) = e^{4W} L_b(\phi,a)$, required to source the given
bulk solution. Because $e^\phi$ and all metric functions behave as
exponentials near the branes --- {\em c.f.}
eq.~\pref{asymptoticpowers} --- the brane action must have the
form $L_b = - \Lambda_b \, e^{\xi_b \phi} \cF \left( a \,
e^{\zeta_b \phi} \right)$, where $\cF(x)$ is an arbitrary function
and the powers $\xi_b$ and $\zeta_b$ are chosen to ensure the
$\eta$-independence in the near-brane regime of
\be
 T_b = - \Lambda_b \, \cW^{4} e^{\xi_b \phi}
 \, \cF \left( a \, e^{\zeta_b \phi} \right) \,,
 \label{sugratensionF}
\ee
for constant $\Lambda_b$. The parameters $\xi_b$ and $\zeta_b$
therefore satisfy
\be
 4\, \omega_b + \xi_b \, q_b =
 \alpha_b + \zeta_b \, q_b = 0 \,.
\ee

In terms of $\cF(x)$, the scalar matching condition becomes
\begin{equation} \label{scalarmatchingNSF}
 q_b = \frac{\kappa^2}{2\pi} \, \left(
  \frac{\pd \, T_b}{\pd\phi} \right)
  = - \frac{\kappa^2 \Lambda_b}{2\pi} \, \cW^4 \, e^{\xi_b \phi}
  \Bigl[ \xi_b \, \cF(x) + \zeta_b \, x\, \cF'(x)
  \Bigr]_{x = a e^{\zeta_b \phi}} \,,
\end{equation}
while the metric matching conditions similarly give
\begin{equation} \label{munumatchingW1F}
 3 \, \omega_b + \alpha_b
 = - \frac{\kappa^2 T_b}{2\pi}
 =  \frac{\kappa^2 \Lambda_b}{2\pi} \, \cW^4 \, e^{\xi_b \phi}
 \cF \left( a \, e^{\zeta_b \phi} \right) \,,
\end{equation}
and so on.

To go further requires making choices for the function $\cF(x)$.
We discuss for simplicity a power-law, $\cF(x) = x^{\sigma_b}$,
which to concretely illustrate the brane-bulk interaction.

\bigskip\noindent{\em Power-law tension: $\cF(x) = x^{\sigma_b}$}

\medskip\noindent
Perhaps the simplest choice for the function
$\cF(x)$ appearing above is a power: $\cF(x) = x^{\sigma_b}$, for
$\sigma_b$ a constant. In this case
\be
 T_b = -\Lambda_b \, \cW^{4} a^{\sigma_b}
 \, e^{\lambda_b \phi} \,, \label{sugratension}
\ee
where $\lambda_b = \xi_b + \zeta_b \sigma_b$, and so
\begin{equation}
 4 \omega_b + \sigma_b \alpha_b + \lambda_b q_b = 0 \,,
\end{equation}
is required to ensure that the $\eta$-dependence cancels in $T_b$
within the near-brane regime. This last equation is to be regarded
as being solved for $\sigma_b$.

The scalar matching condition, eq.~\pref{scalarmatchingNS}, then
boils down to
\be
 q_b = -\lambda_b \, \cW_b^{4} a_b^{\sigma_b} \left(
 \frac{\kappa^2 \Lambda_b}{2\pi} \right) \,. \label{Tpconscondition}
\ee
The $(\mu\nu)$ metric matching condition,
eq.~\pref{munumatchingW}, similarly gives
\be
 3 \, \omega_b + \alpha_b = \cW_b^{4} a_b^{\sigma_b} \left(
 \frac{\kappa^2 \Lambda_b}{2\pi} \right) \label{Tconscondition}
 \,.
\ee
Combining (\ref{Tpconscondition}) and (\ref{Tconscondition}),
gives the parameter $\lambda_b$ as
\begin{equation}
 \lambda_b=-\frac{q_b}{3\omega_b+\alpha_b} \,. \label{powersolution}
\end{equation}
Clearly $q_b < 0$ implies $\lambda_b > 0$ and vice versa, because
$\alpha_b$ and $\omega_b$ are both positive. Notice that
$\lambda_b > 0$ implies $T_b \to 0$ in the `weak-coupling' limit
$e^\phi \to 0$.

Given $\alpha_b$ and $\omega_b$, solving the above conditions
gives $q_b = \pm 2\sqrt{\omega_b (2\alpha_b + 3 \omega_b)}$ (from
eq.~\pref{qconstraint}), $\lambda_b$ (from
eq.~\pref{powersolution}), and the combination $\cW_b^{4}
a_b^{\sigma_b} \left( {\kappa^2 \Lambda_b}/{2\pi} \right)$ (from
eq.~\pref{Tconscondition}). The power of $a$ appearing in $T_b$
works out to be
\be
 \sigma_b  = \frac{4\omega_b}{3\, \omega_b + \alpha_b}
 > 0 \,.
\ee
One might think that the last matching condition, involving $U_b$,
gives an independent equation that can be used to relate
$\omega_b$ to $\alpha_b$, but this turns out not to be independent
due to the relation between $U_b$ and $T_b$ and the constraint,
eq.~\pref{qconstraint}.

\subsubsection*{The 4D perspective}

In this section, we evaluate the full action at its classical
solution to determine the value of $V_{\text{eff}}$ at its
minimum. For supergravity the full bulk action evaluates to a
total derivative at any classical solution, giving
\begin{equation}
 S_{\ssB,\text{ext}} = \frac1{2\kappa^2} \int \exd^6x
 \, \sqrt{-g} \; \Box\phi = \frac{\pi}{\kappa^2}
 \int \exd^4x \, \sqrt{-\hat g} \;
 \Bigl[ \pd_\eta\phi \Bigr]^{\infty}_{-\infty}
 = -\sum_b\frac{T_b'}2 \,.
\end{equation}
Adding to this the brane action and Gibbons-Hawking term, which
combine to
\begin{equation}
 \sum_b \Bigl( S_{GH}+S_b \Bigr) = - \int d^4x
 \sqrt{\hat g} \; U_b
\end{equation}
gives the total action evaluated at the classical solution
\begin{equation}
 S_{\text{ext}} = -\int \exd^4x \sqrt{-\hat g} \;
 \sum_b \Bigl( U_b + \frac{T_b'}{2} \Bigr)
 \,.
\end{equation}
Comparing this with eq.~\pref{Sextgeneral} (for $n=6$) gives
\be
 V_{\rm\,eff}(\phi_0) = -
 \sum_b \left( U_b + \frac{T_b'}{2} \right) \,.
\ee

Using this in the four-dimensional Einstein equations gives the 4D
curvature
\begin{equation}
 \hat R = -12H^2 = -4\kappa_\ssN^2 V_{\text{eff}}(\phi_0) \,,
\end{equation}
and so
\begin{equation}
 H^2 = \frac{\kappa_\ssN^2}{3} \, V_{\text{eff}}
 = - \frac{\kappa_\ssN^2}{3} \sum_b \left(
 U_b + \frac{T_b'}{2} \right)
 = - \frac{2\pi\kappa_\ssN^2}{3\kappa^2}
 \sum_b \left( \omega_b + \frac{q_b}2 \right),
\end{equation}
where the last equality uses the matching conditions to rewrite
$U_b$ and $T'_b$ in terms of the bulk solution. This agrees with
the bulk field equations, eq.\pref{IntegratedCBA-GGP}, and so
shows that the $4D$ and $6D$ pictures agree. In order to identify
the value of $\phi_0$ itself requires calculating $V_{\text{eff}}$
away from its minimum, which requires a full dimensional reduction
of the supergravity action.

\section{Conclusions}

This paper summarizes the bulk-brane matching conditions for
codimension-2 objects (following the presentation given for
scalar-tensor theories in \cite{cod2matching}, with
generalizations to include a general coupling to the Maxwell field
\cite{uvcaps}), and describes several applications to
higher-dimensional brane systems: F-theory compactifications
involving space-filling codimension-2 D7-branes situated within 10
dimensions; unwarped 3-brane flux compactifications in 6
dimensional scalar-Maxwell-Einstein theory; and warped and
unwarped 3-brane flux compactifications of 6D chiral gauged
supergravity. The latter two cases involve geometries that are
maximally symmetric --- but possibly curved --- in the directions
parallel to the branes.

The comparison with the F-theory compactifications provides a
sanity check on the junction conditions, since both the brane and
bulk actions are explicitly known for Type IIB string vacua
\cite{polchinski}, as are explicit solutions for the surrounding
bulk geometry \cite{scs}. We show that the near-brane asymptotic
form of the bulk configurations in this case precisely agrees with
what the matching conditions would predict, given the explicit
D7-brane action. Furthermore, this comparison lies within the
weak-coupling regime since the bulk solution implies the string
coupling becomes weak in the near-brane limit.

When applied to six-dimensional systems, the bulk-brane matching
conditions can provide a stabilization mechanism for the bulk
scalars (like a bulk axion, or the dilaton) provided the brane
couplings break the appropriate symmetry that protects the
scalar's mass. When this is so, the value to which the scalar
stabilizes can be understood from the higher-dimensional point of
view as being due to the consistency of the matching conditions at
the two branes. Alternatively it can be regarded as the value
which minimizes the effective potential in the low-energy,
on-brane action below the KK scale, although this requires a
calculation of the potential away from its minimum.

Although many of the bulk solutions considered in six dimensions
(supersymmetric or not) have de Sitter curvature along the four
brane directions \cite{conical,scalingpaper},
we show that for 6D gauged chiral supergravity
only 4D-flat branes can be sourced by positive-tension branes. To
establish this we first show that for any 6D theory a
codimension-2 brane tension must be negative whenever the warp
factor tends to zero near the brane. We then prove that the
supergravity field equations imply the warping vanishes near the
brane unless the near-brane geometry has a conical singularity.
Finally, the desired result follows once the field equations are
used to see that any geometry having only conical singularities
necessarily is flat in the 4 brane directions.

This necessity for negative tension in order to obtain de Sitter
and anti-de Sitter branes echoes the various no-go theorems for
finding 4D-de Sitter solutions from extra-dimensional gravity
\cite{nogos}, even though the curvatures of the bulk geometries
considered make these theorems not directly apply. This suggests
that the curvature assumptions made in these theorems may be
somewhat stronger than is necessary.

The relation to 4D de Sitter geometries has potential applications
to searches for cosmic inflation within an extra-dimensional
context. This is because inflationary configurations often lay
nearby pure de Sitter solutions. In particular, a broad class of
time-dependent solutions are known \cite{scalingpaper} for the
bulk field equations in 6D supergravity, and for some of these the
on-brane 4D geometry is likely to undergo an accelerated
expansion. The extension of the arguments of this paper to these
time-dependent situations would be most worthwhile, since they
could provide instances of explicit inflationary models for which
there is both a higher- and lower-dimensional understanding of why
the universe accelerates. (By contrast, current inflationary
models typically rely on the low-energy 4D effective theory to
conclude that the universe inflates.) Work along these lines is in
progress \cite{later}.

\section*{Acknowledgements}

We wish to thank Hyun-Min Lee and A. Papazoglou for discussions
about codimension-2 branes and inflation, as well as Anshuman
Maharana and Fernando Quevedo for advice about F-theory. Our
research has been supported in part by funds from the Natural
Sciences and Engineering Research Council (NSERC) of Canada,
McMaster University and Perimeter Institute. Research at the
Perimeter Institute is supported in part by the Government of
Canada through NSERC and by the Province of Ontario through the
Ministry of Research and Information (MRI).

\end{document}